\documentclass[pra,aps,amsmath,amssymb,twocolumn,longbibliography]{revtex4-1}
\usepackage{graphicx,verbatim}
\usepackage{color}
\usepackage{ulem}
\newtheorem{thm}{Theorem}

\newcommand{\halv}{\frac{1}{2}}
\newcommand{\be}{\begin{equation}}
\newcommand{\ee}{\end{equation}}

\newcommand{\bex}{\begin{eqnarray}}
\newcommand{\eex}{\end{eqnarray}}
\newcommand{\bmin}{\begin{center}\begin{minipage}{460pt}}
\newcommand{\emin}{\end{minipage}\end{center}}

\newcommand{\betaf}{\frac{\beta}{2}}
\newcommand{\lf}{\frac{1}{2}l}
\newcommand{\alfaf}{\frac{\alpha}{2}}
\begin{document}

\title{Extending  quantum  mechanics   entails  extending special
  relativity}   
\author{S.    Aravinda}   \email{aru@poornaprajna.org}
\author{R.            Srikanth}          \email{srik@poornaprajna.org}
\affiliation{Poornaprajna    Institute    of   Scientific    Research,
  Sadashivnagar, Bangalore, India}

\begin{abstract}
The   complementarity  between   signaling  and   randomness  in   any
communicated  resource   that  can  simulate  singlet   statistics  is
generalized by relaxing  the assumption of free will in  the choice of
measurement  settings.   We  show  how  to  construct  an  ontological
extension  for quantum  mechanics (QM)  through the  \textit{oblivious
  embedding} of a sound simulation  protocol in a Newtonian spacetime.
Minkowski or other intermediate spacetimes  are ruled out as the locus
of  the embedding  by virtue  of hidden  influence inequalities.   The
complementarity transferred from a simulation to the extension unifies
a  number  of results  about  quantum  nonlocality, and  implies  that
special  relativity   (SR)  has  a  different   significance  for  the
ontological model and for the  operational theory it reproduces.  Only
the latter, being experimentally accessible, is required to be Lorentz
covariant.  There may be certain Lorentz non-covariant elements at the
ontological level,  but they will  be inaccessible at  the operational
level   in  a   valid  extension.    Certain  arguments   against  the
extendability of QM,  due to Conway and Kochen (2009)  and Colbeck and
Renner (2012), are  attributed to their assumption  that the spacetime
at the ontological level has Minkowski causal structure.
\end{abstract}
\maketitle

\section{Introduction\label{sec:intro}}   

Bell's theorem proves that quantum nonlocality cannot be reproduced by
any  local  realistic model \cite{Bel64}.  A  simple instance of
nonlocality  is demonstrated  by  the quantum  violation  of the  CHSH
inequality \cite{CHSH}
\begin{equation}
\Lambda(\textbf{P}) = \sum_{a,b} (-1)^{a\overline{b}} E(a,b) \le 2,
\label{eq:bi}
\end{equation}
with inputs  $a, b  \in \{0,1\}$  and the  expectation value  $ E(a,b)
\equiv P(x=y|ab) - P(x\ne y|ab)$, where the respective outputs are $x,
y \in  \{0,1\}$. Here  $\overline{b}\equiv b\oplus 1$,  where $\oplus$
indicates  addition   modulo  2.    Bell's   theorem  places  no
  restriction on how  much must be given up in  a hidden variable (HV)
  model or ``extension''.  Bell's result has been strengthened by the
relaxing  of localism  \cite{Leg03,GPK+07},  and further  strengthened
recently  through  the  ruling  out of  any  (nontrivial)  local  part
\cite{BKP06,BBG+08,CR08}.

The above works entail that any  HV model of singlet correlations must
be entirely  nonlocal.  Two recent works,  invoking special relativity
(SR) and a certain version of  free will, present arguments that would
also rule out deterministic  nonlocal extensions \cite{CK06,*CK09} and
indeterministic nonlocal extensions also  \cite{CR11}.  On this basis,
they   argued   that   nonlocal   extensions  of   QM   like   Bohmian
\cite{Boh52a,*Boh52b}  and   GRW  collapse  models   \cite{GRW86}  are
incompatible with with SR and free will.

These conclusions were contested \cite{BG07,Tum07,GTT+10,Sua10} on the
basis of  two broad grounds:  (a) that Bohmian and  GRW collapse
  theories  are already  known to  be observationally  compatible with
  SR. In  the case  of the  former, this has  been known  since Bohm's
  original  works  and also  from  Bell's  writings \cite{Bel87}.   As
  regards    their   ontological    compatibility    with   SR,    cf.
  \cite{Sua10,Tum06a,*Tum06b,DGN+14,BDG+14}.  Thus  these models have
no  obvious  obstacle  to  admitting   free  will;  (b)  that  FW  and
no-signaling are  logically independent, so  that invoking FW  to rule
out predictively superior extensions is untenable \cite{GR0,GR00}.

This conflict illustrates that the  ``tension'' between SR and quantum
nonlocality  is still  not unequivocally  resolved. In  this work,  we
provide a  resolution to  this conflict  by identifying  the different
assumptions behind the conflicting claims.


Our  approach to  the resolution  will be  through the  following four
steps, an  expansion of which  is given  in the overview  presented in
Section  \ref{sec:over}.   In  the  first step,  we  clarify  (Section
\ref{sec:defn})  that nonlocal  correlations must  be viewed  from two
levels  or  layers:  the   \textit{operational}  level  accessible  to
experimentalists and  the \textit{ontological} level, where  the HV's,
by definition inaccessible and  unknowable, live \cite{Spe05,CW12}. In
particular, we define  unpredictability and (operational) no-signaling
as operational concepts, having indeterminism and \textit{ontological}
no-signaling   as   their   ontological  counterparts.    In   Section
\ref{sec:spontan},  we define  free will  as appropriate  to Bell-type
experiments, and introduce the concept of \textit{spontaneity}, as the
operational equivalent  of free  will. We clarify  in this  first step
that   only   operational   concepts,  and   not   their   ontological
counterparts,  are  required  to  be  Lorentz  covariant,  since  only
operational   quantities    are   experimentally    accessible.    For
terminological    clarity,     we    propose    the     concepts    of
\textit{randomness},  \textit{signaling} and  \textit{freedom} as  the
level-neutral  counterparts   of  the   three  operational/ontological
concepts discussed above.

As  the  second step  in  our  argument,  we  present a  protocol  for
simulating   singlet    statistics   (Section    \ref{sec:simu}).    A
complementarity of the  signaling and randomness of  resources used in
the protocol, as modified by relaxing  the assumption of free will, is
formulated in Section \ref{sec:srf}.

As the  third step,  we present in  Section \ref{sec:srx}  an explicit
procedure  to  convert any  sound  simulation  protocol into  a  valid
ontological extension of QM.  There  is a two-fold subtlety about this
conversion: (a) The protocol must be  embedded in a Newtonian (and not
Minkowski)  spacetime; (b)  The embedding  must be  oblivious, meaning
that certain simulation parameters will map to the ontological theory,
and hence must be unknowable to physical observers, Alice and Bob.

This exercise  will allow us  to clarify that  an extension of  QM may
contain ontological features that  are Lorentz non-covariant. However,
these will either be suitably averaged out or physically inaccessible,
such  that  the  resulting  operational theory  will  conform  to  SR,
assuming the soundness of the simulation protocol.

As  the  fourth and  last  step  in  our  approach, we  show  (Section
\ref{sec:XSIC})   how    the   freewill-relaxed   randomness-signaling
complementarity   of  simulation   resources  carries   over  to   the
ontological  extension under  the embedding  procedure. All  the above
stronger forms of Bell inequalities will be derived as consequences of
this   transplanted  complementarity   in  the   context  of   singlet
statistics.  Here  we will  finally be  in a  position to  revisit and
unpack the  above mentioned  debate in  the literature  regarding free
will, no-signaling and unextendability of  QM.  We conclude in Section
\ref{sec:conclu}.

\section{Overview of results \label{sec:over}}

In  view of  the fact  that  the problem  dealt with  is fraught  with
conceptual difficulties, we present in  this section an outline of the
arguments  in this  work. This  outline essentially  expands the  four
steps mentioned above.

\subsection{Definitions: Operational and ontological levels}

It  is known  that two  levels of  description come  into play  in the
description of nonlocality \cite{Spe05,CW12}: the \textit{operational}
or observational level and the \textit{ontological} or HV level.  Here
we wish to stress that these two levels are constrained differently by
relativistic causality.  For our  purpose, it suffices to characterize
the correlations in  terms of three concepts or  resources, which must
be defined at  both the operational and ontological  levels. The three
concepts are  randomness, (no-)signaling  and freedom  (in measurement
settings) in  a correlation, all three  of which must be  specified at
the operational  and ontological levels.  Their  definitions are given
in Sections \ref{sec:defn} and \ref{sec:spontan}.

We list  the operational  and ontological  equivalents of  these three
concepts in Table \ref{tab:oplist}.
\begin{table}
\begin{tabular}{c|c|c}
\hline
\textbf{Level-neutral} & \textbf{Operational} & \textbf{Ontological} \\
\textbf{Concept} & \textbf{quantity} & \textbf{counterpart} \\
\hline
\textit{Randomness} & unpredictability & indeterminism 
\\ \hline
\textit{No-signaling} & operational  &  ontological \\
  & no-signaling & no-signaling \\ \hline
\textit{Freedom} & spontaneity  & free will \\ \hline
\end{tabular}
\caption{Operational  and ontological  equivalents in  the context  of
  quantum  nonlocal  correlations.   Only operational  quantities  are
  required to be Lorentz covariant.}
\label{tab:oplist}
\end{table}
The concepts ontological no-signaling  and operational no-signaling in
Table  \ref{tab:oplist} are  close  to the  concepts  of locality  and
signaling locality in Ref. \cite{CW12},  and similarly the concepts of
unpredictability and indeterminism  in Table \ref{tab:oplist} parallel
the  like-named  concepts  in   Ref.   \cite{CW12}.   The  concept  of
spontaneity is introduced  here is the operational  equivalent of free
will \cite{Hal10}.

In each case,  the operational quantity is obtained  by averaging over
the  ontic  or underlying  state  $\lambda$  of  the system.   As  the
experimentally accessible  variables, only the  operational quantities
in  Table \ref{tab:oplist}  are required  to amenable  to a  covariant
description.  By contrast,  ontological  variables  are by  definition
unknowable  and  inaccessible, and  thus  they  are not  compelled  by
relativistic causality to satisfy Lorentz covariance.  For example, an
extension  may violate  ontological  no-signaling, but  this will  not
matter  provided  the  theory  reproduced  on  the  operational  level
satisfies operational no-signaling.

\subsection{Result 1: Freedom-relaxed complementarity
\label{sec:res1}}

It  will be  convenient  to  think of  a  protocol $\mathfrak{S}$  for
simulating  nonlocality  (in the  context  of  singlet statistics)  as
two-layered:  the  \textit{base layer}  consisting  of  the input  and
output  random variables  $A, B,  X$ and  $Y$ of  Alice and  Bob in  a
practical experiment;  and the  \textit{meta-layer} consisting  of the
classical randomness (denoted $\chi$ and $\chi^+$ here) that Alice and
Bob pre-share  in the simlation,  and the resource  $\mathcal{R}$ that
Alice communicates  to Bob during  the simulation run.   In accordance
with this  two-layering, it will  sometimes be convenient to  refer to
$A, B,  X, Y$ as  the \textit{base  data}, and to  $\mathcal{R}, \chi,
\chi^+$ as the \textit{meta data}.

It is  known that the  resource $\mathcal{R}$ consisting of  a one-bit
signal  \cite{TB03}   or  of   a  single  Popescu-Rohrlich   (PR)  box
\cite{CGM+05}  (which   is  non-signaling   but  with   maximal  local
randomness) suffices to simulate the nonlocal statistics of a singlet.
More  generally,  the  communicated  resource  $\mathcal{R}$  shows  a
\textit{complementarity} between signaling  $S_R$ and local randomness
$I_R$ \cite{Hal10b, KGB+11, ASQIC}.  Our first result is to generalize
this complementarity by relaxing the assumption of experimenters' free
will $F$ \cite{Hal10}.

Reducing freedom $F$ relaxes the above complementaristic constraint on
$S_R$ and $I_R$, until (at $F=\frac{2}{3}$) there is no bound on these
two  quantities, i.e.,  the  correlations can  be  reproduced using  a
mixture of  local-deterministic correlations.  Under  maximal freedom,
the complementarity has the form:
\begin{equation}
S_R + 2I_R \ge 1
\label{eq:1comp}
\end{equation}
for  singlet simulation,  a  result proved  in Section  \ref{sec:srf}.
Complementarity (\ref{eq:1comp}) implies that  under the assumption of
maximal freedom, the amount of randomness and signaling cannot both be
arbitrarily low in resource $\mathcal{R}$ that is suitable to simulate
singlet statistics.


\subsection{Result 2: Elevating a simulation protocol
to an ontological model via Oblivious embedding in a Newtonian 
spacetime \label{sec:res2}}

We show how to elevate protocol $\mathfrak{S}$ to an ontological model
by  letting ``Nature  run $\mathfrak{S}$''  in spacetime.   This seems
intuitively  clear, but  it would  appear that  communicating resource
$\mathcal{R}$  in  spacetime  would  violate  no-signaling  and  hence
relativistic   causality   in   the  case   of   spacelike   separated
measurements.  Key to  seeing that this is not so,  is to observe that
precisely the base  layer of protocol $\mathfrak{S}$ is  mapped to the
operational  theory,  whereas  the  meta  layer  gets  mapped  to  the
ontological  model.  Accordingly,  we  have the  following recipe  for
mapping of data in $\mathfrak{S}$ to variables in spacetime:
\begin{eqnarray}
{\rm    Base~data~of~}\mathfrak{S}  &\longrightarrow&    {\rm
  operational~variables}    \nonumber\\   {\rm   Meta~data~of~}
\mathfrak{S}
   &\longrightarrow&  {\rm   ontological~
  variables}
\label{eq:basemeta}
\end{eqnarray}
Therefore, the type of  no-signaling violated during the communication
of $\mathcal{R}$ is \textit{ontological}.  Since ontological variables
are by definition inaccessible to Alice  and Bob (the observers in the
operational theory), no violation  of relativistic causality occurs at
the operational level.  On the other  hand, the base data by itself is
consistent with  no-signaling (in that mutual  information $I(A:Y)=0$,
etc.),  and thus  there  is  no difficulty  in  mapping  this data  to
operational variables.

Obviously, the spacetime in which  ontological elements live cannot be
governed  by  SR,  but  instead  should  be  governed  by  a  suitable
``ontological extension of  SR'' (SRX).  A SRX is  a relativity theory
in  which the  causal  structure  of SR  is  replaced  by another  one
(appropriate for  quantum nonlocal phenomena)  in which at  each event
\textbf{e},  the causally  connected region  strictly encompasses  the
light cones of SR.  (Other details of SR are not germane here.)  A SRX
is  a  kind  of  $v$-causal  model   of  the  type  proposed  in  Ref.
\cite{SBB+08}.     Within   this  framework   (of  deriving   an
  ontological model  from a simulation  protocol), it turns  out that
the only  allowed SRX  is the one  equipped with  Galilean invariance,
i.e.,  the Newtonian  spacetime, because  for any  other SRX,  one can
always  produce an  experimental  configuration that  would predict  a
breakdown in the quantum  correlations.  Hidden influence inequalities
\cite{SBB+08} can be  constructed that make use of  this breakdown for
superluminal signaling at the operational level.

 This conclusion does not automatically apply to extensions of QM
  that cannot be analyzed  manifestly as simulation protocols embedded
  in spacetime, i.e., in the  pattern of Eq.  (\ref{eq:basemeta}).  In
  such  cases,  Lorentz  covariance   may  indeed  hold  for  elements
  recognized   as   ontological   in   the   extension,   cf.    Refs.
  \cite{Sua10,Tum06a,*Tum06b,DGN+14,BDG+14} as regards GRW and Bohmian
  models.  In  this light,  our result can  broadly be  interpreted as
  showing that for  any model of a nonlocal theory  like QM, under the
  assumption of free  will, there would be  fundamental influences and
  fundamental correlations not conforming to the lightcone structure.

We thereby have a procedure  to elevate any protocol $\mathfrak{S}$ to
an  ontological  extension  of  (a   fragment  of)  QM,  by  embedding
$\mathfrak{S}$ in a  SRX, as described above.  This shows  in a simple
way that extension of this type will contain some ontological elements
that aren't Lorentz covariant.  But  the physical agents Alice and Bob
will be oblivious to (i.e., unable to access) them.  Thus, conformance
to no-signaling is automatically guaranteed at the operational level.

\subsection{Discussion: 
SR and complementarity in the derived extension}

The signaling and randomness in the meta data $\mathcal{R}$ (discussed
above in  Section \ref{sec:res1})  is transferred from  the simulation
scenario  to  the  spacetime   scenario  under  the  embedding  scheme
(discussed  in  Section  \ref{sec:res2}).   By  virtue  of  assignment
(\ref{eq:basemeta}), we then have at the ontological level:
\begin{eqnarray}
S_R &\rightarrow& S_\lambda \nonumber\\
I_R &\rightarrow& I_\lambda.
\label{eq:conversion}
\end{eqnarray}
With this identification, Eq.  (\ref{eq:1comp}) becomes:
\begin{equation}
S_\lambda + 2I_\lambda \ge 1,
\label{eq:2comp}
\end{equation}
i.e.,   a   complementarity    between   ontological   signaling   and
indeterminism. The  is a stronger  form of Bell's theorem,  which only
says $S_\lambda  + 2I_\lambda >  0$.  By identifying meta  data $\chi,
\chi^+$ with the underlying ontological state, we can identify freedom
$F$ with free  will (and will use  the same symbol, since  there is no
confusion).   The  complementarity  (\ref{eq:2comp}) can  be  used  to
obtain equivalent  derivations of the  various stronger forms  of Bell
inequalities            mentioned           earlier,            namely
\cite{Leg03,GPK+07,BKP06,BBG+08,CR08},  as  well as  the  mathematical
essence of Refs.  \cite{CK06,*CK09} and \cite{CR11}.

The operational theory can be  considered as a ``trivial ontology'' by
setting the signaling $S_\lambda^{\rm triv}  \equiv S = 0$, from which
and Eq. (\ref{eq:2comp}) it  follows that $I^{\rm triv}_\lambda \equiv
I  = \halv$.   In other  words,  the operational  theory must  contain
maximum  unpredictability.   Ontologically,  we  can  have  predictive
superiority,    i.e.,   $I_\lambda    <   \halv$,    and   from    Eq.
(\ref{eq:2comp}),  we   find  $S_\lambda>0$.   This  means   that  any
predictively  superior   extension  will  contain  signaling   at  the
ontological level, which can  (as indicated in Section \ref{sec:res2})
coexist peacefully with no-signaling at the operational level.


We stress that  this conclusion was already reached  by the proponants
of Bohmian mechanics and GRW models  with respect to their own models.
What  is new  to  our work  is  to  identify   for  the class  of
  ontological  models  based  on  a protocol  for  simulating  quantum
  nonlocality, which elements in  a predictively superior extension of
  QM  are  necessarily Lorentz-covariant  and  which  elements may  be
  non-covariant:    namely,   the    operational   and    ontological,
  respectively.  As  one particular  application of this  result, our
work gives a general and simple explanation of why the technical no-go
results of  the type  \cite{CK06,*CK09,CR11} cannot be  interpreted as
prohibiting  such extensions  on  grounds  of relativity,  essentially
because  the  non-covariance  that   they  identify  pertains  to  the
ontological elements  for this class of extensions.

\section{Resources in operational 
and ontological theories: signaling and randomness\label{sec:defn}}

By  an  ``operational  theory''  we mean  a  theory  characterized  by
physical  measurements  and  observations  by  one  or  more  parties,
outcomes   and  the   corresponding   conditional  correlations.    An
operational   theory  may   contain  counterintuitive   features  like
non-signaling nonlocality, for which an ``ontological model'', such as
a HV theory,  attempts to provide a more  intuitive and classical-like
explanation  using  variables  that  may not  be  directly  accessible
physically.

\subsection{Signaling: operational and ontological}

A  bipartite  correlation  $\textbf{P}\equiv P_{XY|AB}$  generated  by
measurements in an operational theory, is non-signaling if
\begin{eqnarray}
P_{X|AB}  &=& P_{X|A},\nonumber\\
P_{Y|AB}  &=& P_{Y|B}.
\label{eq:opnosig}
\end{eqnarray}
where $A$  and $X$  (resp., $B$  and $Y$)  are Alice's  (resp., Bob's)
input and  output spacetime-labelled random variables  (abbreviated to
SVs).  By  relativity considerations, Eq.  (\ref{eq:opnosig}) must
hold if $A$ and $B$ are spacelike separated and freely chosen.

Now  suppose   we  extend  conditions   (\ref{eq:opnosig}),  requiring
no-signaling additionally in a HV theory.  Then we may require:
\begin{eqnarray}
P_{X|AB\lambda} & =& P_{X|A\lambda},\nonumber\\
P_{Y|AB\lambda}  &=& P_{Y|B\lambda},
\label{eq:ontnosig}
\end{eqnarray}
where $\lambda$ is the HV describing the ontic state in the underlying
ontological  theory.  Eq.   (\ref{eq:ontnosig})  is a  version of  the
\textit{ontological}  no-signaling condition.   If  Alice and  Bob
choose their measurement settings freely, then it is not necessary for
this condition to  be satisfied.  This point is crucial  here, in that
the nay-sayers  in the above debate  treat it at par  with operational
no-signaling.

Instead, what  is necessary by virtue  of requiring SR to  hold in the
operational theory, is the following: If $\rho(\lambda|AB)$ represents
the probability distribution of $\lambda$ conditioned on the inputs in
the operational  theory, then  we require  that $P_{X|AB}  \equiv \int
\rho(\lambda|AB)  P_{X|AB\lambda}d\lambda$ and  $P_{Y|AB} \equiv  \int
\rho(\lambda|AB)    P_{Y|AB\lambda}d\lambda$   should    satisfy   the
\textit{operational}  no-signaling conditions  (\ref{eq:opnosig}),
if    Alice     and    Bob     have    full    free     will,    i.e.,
$\rho(\lambda|AB)=\rho(\lambda)$.

As  we  show  later,  the violation  of  ontological  no-signaling  is
\textit{necessary} for non-trivial extensions of QM.  Recognizing this
beneficial aspect  of ontological  signaling is  key to  resolving the
aforementioned debate.
\begin{center}
\begin{table} 
\begin{tabular}{c||c|c|c|c|c|c|c|c}
\hline  $ab$  &  $\textbf{d}^{0_1}$ & $\textbf{d}^{1_1}$ &
$\textbf{d}^{2_1}$ & $\textbf{d}^{3_1}$ & $\textbf{d}^{4_1}$  & $\textbf{d}^{5_1}$ & $\textbf{d}^{6_1}$ & $\textbf{d}^{7_1}$
\\ \hline 
00 & 00 & 11 & 00 & 11 & 00 & 11 & 00 & 11 \\ 
01 & 00 & 11 & 00 & 11 & 00 & 00 & 11 & 11 \\ 
10 & 01 & 01 & 10 & 10 & 10 & 01 & 10 & 01 \\ 
11 & 00 & 00 & 11 & 11 & 00 & 00 & 11 & 11 \\ 
\hline
\end{tabular}
\caption{The    complete   list    of   deterministic    1-bit   boxes
  $\textbf{d}^{j_1}$   that  violate   Ineq.   (\ref{eq:bi})   to  its
  algebraic  maximum of  +4.   Every 1-bit  box $\textbf{d}^{j_1}$  is
  paired   with  an   ``anti-box'',   e.g.,  $\textbf{d}^{0_1}$   with
  $\textbf{d}^{3_1}$, with  complementary outputs.  Together  with the
  deterministic correlations of  Table \ref{tab:bm0}, these constitute
  the  full  set   of  extreme  points  for   the  signaling  polytope
  $\mathfrak{F}^P$.}
\label{tab:bm1}
\end{table} 
\end{center}

If the  condition for  operational no-signaling  (\ref{eq:opnosig}) or
for  ontotological no-signaling  (\ref{eq:ontnosig}) is  violated, the
resultant operational  signal or ontological signal  can be quantified
in  a  variety  of  ways.  One   such  is  described  below.   In  the
two-input-two-output case, the  \textit{operational} signal from Alice
to Bob  ($S^{A\rightarrow B}  $) and Bob  to Alice  ($ S^{B\rightarrow
  A}$) can be quantified as
\begin{eqnarray}
S^{A\rightarrow B}  &=&
\sup_{b}|P_{y|0,b}   -   P_{y|1,b}| \nonumber\\ 
S^{B\rightarrow   A}   &=&
\sup_{a}|P_{x|a,0}   -   P_{x|a,1}|,
\end{eqnarray}
where  $P_{x|a,b}$  (resp.,  $P_{y|a,b}$) is  Alice's  (resp.   Bob's)
marginal distribution.  The operational signal is quantified as
\begin{equation}
S = \max\{S^{A\rightarrow B}, S^{B\rightarrow A}\},
\label{eq:s()}
\end{equation}   
with the condition $S=0$  implying (\ref{eq:opnosig}).  In general, $0
\le S \le 1$ and specifically $S\le C$, where $C$ is the communication
cost  of correlation  \textbf{P}  \cite{ASIQSA}.   If the  operational
no-signaling condition (\ref{eq:opnosig}) is violated, then $S>0$.

In  analogy with  definition (\ref{eq:s()}),  the \textit{ontological}
signal  $S_\lambda$  can be  quantified  by  replacing $P_{x|ab}$  and
$P_{y|ab}$    in     (\ref{eq:s()})    by     $P_{x|ab\lambda}$    and
$P_{y|ab\lambda}$,  where the  latter two  quantities are  Alice's and
Bob's  marginal  distribution  in  the  ontic  state  $\lambda$.   The
condition $S_\lambda=0$ implies the ontological no-signaling condition
(\ref{eq:ontnosig}).    In   general,   $0  \le   S_\lambda   \le   1$
\cite{ASIQSA}.  We only require that $S=0$.  If the operational theory
contains correlation  \textbf{P} given by  a PR box  \cite{PR94}, then
$S_\lambda=1$ and $S=0$.

As  an illustration  of ontological  signaling leading  to operational
no-signaling, consider the determinsitic  distributions given in Table
\ref{tab:bm1} as ontic  states $\lambda$ in an  underlying theory.  As
each of them requires 1 bit  of communication to be simulated, we will
refer  to   them  as   ``1-bit  boxes'',   denoted  $\textbf{d}^{j_1}$
($j=0,1,2\cdots   ,7$).     

An example  of a  1-bit box is  $\textbf{d}^{0_1}$, which  denotes the
probability   distribution    $d^{0_1}_{xy|ab}   \equiv   \delta^0_{x}
\delta_{y}^{a\cdot(b\oplus1)}$.   We  can obtain  the  PR  box at  the
operational      level     by      uniformly     mixing      ``boxes''
$\lambda=\textbf{d}^{0_1}$  and  $\lambda=\textbf{d}^{3_1}$.  In  this
case,  there is  an  ontological  signal from  Alice  to  Bob in  both
individual  cases of  $\lambda$ in  violation of  (\ref{eq:ontnosig}).
However,  in  the  PR  box  realized at  the  operational  level,  the
operational  no-signaling condition  (\ref{eq:opnosig}) is  satisfied.
Conversely,  ontologically  non-signaling  ontic states  can  lead  to
operational signaling, when  the free will of Alice or  Bob is reduced
through  non-trivial  $\rho(\lambda|AB)$,   as  discussed  in  Section
\ref{sec:spontan}.

\subsection{Unpredictability and indeterminism \label{sec:rand}}

We denote by  $I$ the degree of unpredictability,  or local randomness
in  the  operational  theory,   generated  by  measurement  of  either
observer:
\begin{equation}
I \equiv \sup_{a,b} \min_{z} \{P_{z|a,b}\},
\label{eq:I}
\end{equation}
where $z$ is the outcome on  any one of the parties \cite{Hal10b}.  In
general, $0  \le I  \le \frac{1}{2}$.  The ontological  counterpart of
$I$,  which   is  indeterminism,   denoted  $I_\lambda$,   is  defined
analogously, with additional conditioning on $\lambda$:
\begin{equation}
I_\lambda \equiv \sup_{a,b} \min_{z} \{P_{z|a,b,\lambda}\}.
\label{eq:Ilambda}
\end{equation}
A deterministic system  is predictable, but the converse  is not true.
For   the  model   for  the   PR  box   mentioned  earlier,   we  find
$I=\frac{1}{2}$  and $I_\lambda=0$.   Thus the  operational theory  is
maximally   unpredictable,  but   the  underlying   theory  is   fully
deterministic.

\section{Free will and spontaneity\label{sec:spontan}}

The question  of what free will  is, and whether it  exists in Nature,
has  been debated  for centuries  in philosophy  \cite{HS15}.  In  the
context  of Bell  tests,  free will  is  taken to  be  the freedom  or
uncorrelatedness of the observers' choice of measurement settings from
factors lying  to the  past.  Two  relevant and  sometimes contentious
questions here are: What factors to be free from? What is the scope of
the  past? The  answers depend  on the  type of  freedom in  question:
whether it  is operational  or ontological.  Following  convention, we
identify the ontological variety with the term ``free will''.

Ref.  \cite{Hal10} defines  free will $F$ as a measure  of Alice's and
Bob's choices being uncorrelated with the underlying state $\lambda$:
\begin{equation}
F \equiv  1 - \frac{1}{2}\left(\sup_{a,  a^\prime , b,  b^\prime} \int
d\lambda |\rho(\lambda|a,b) - \rho(\lambda|a^\prime,b^\prime)|\right),
\label{defn:Hal10}
\end{equation}
where $\rho(\lambda|a,b)$ is the probability distribution of $\lambda$
conditioned on  input $a,b$.   Free will  so quantified  satisfies the
bound $0 \le  F \le 1$.  

Even with  a reduction of free  will by a fraction  $\frac{1}{3}$, the
CHSH  inequality can  be violated  to  the algebraic  maximum using  a
local-deterministic  model. For  the  eight local-deterministic  boxes
with $\Lambda=+2$, given in Table \ref{tab:bm0}, 
this is  proven below.

As  the  correlations in  Table  \ref{tab:bm0}  require zero  bits  of
communication  to be  simulated, we  shall  refer to  them as  ``0-bit
boxes'', and denote them  by $\textbf{d}^{j_0}$ ($j=0,1,2\cdots$).  An
example for  a 0-bit box  is $\textbf{d}^{2_0}$, which  represents the
probability        distribution         $d^{2_0}_{xy|ab}        \equiv
\delta^0_{x}\delta^{b\oplus1}_{y}$,  where   $\delta^\mu_\nu$  is  the
Dirac delta function.
\begin{center}
\begin{table} 
\begin{tabular}{c||c|c|c|c|c|c|c|c}
\hline  $ab$  &  $\textbf{d}^{0_0}$  &  $\textbf{d}^{1_0}$ &  $\textbf{d}^{2_0}$  &  $\textbf{d}^{3_0}$  &
$\textbf{d}^{4_0}$& $\textbf{d}^{5_0}$ & $\textbf{d}^{6_0}$ & $\textbf{d}^{7_0}$
\\ \hline 
00 & 00 & 00 & 01 & 11 &  00 & 10 & 11 & 11 \\ 
01 & 00 & 00 & 00 & 10 &  01 & 11 & 11 & 11 \\ 
10 & 00 & 10 & 01 & 01 &  10 & 10 & 01 & 11 \\ 
11 & 00 & 10 & 00 & 00 &  11 & 11 & 01 & 11 \\ 
\hline
\end{tabular}
\caption{The    complete   list    of   deterministic    0-bit   boxes
  $\textbf{d}^{j_0}$,    for     which    $\Lambda=+2$     in    Ineq.
  (\ref{eq:bi}). The  first column lists  the inputs, while  the other
  columns give the outputs corresponding to the box.}
\label{tab:bm0}
\end{table} 
\end{center}

We consider a method for reduction  of free will effected by requiring
that   Alice's   and  Bob's   choice   of   inputs  will   depend   on
$\lambda=\textbf{d}^{j_0}$ according to:
\begin{equation}
\rho(ab|\lambda)=\begin{matrix}
\textbf{d}^{0_0}  &  \textbf{d}^{1_0} &  \textbf{d}^{2_0}  &  \textbf{d}^{3_0}  &
\textbf{d}^{4_0} & \textbf{d}^{5_0} & \textbf{d}^{6_0} & \textbf{d}^{7_0}
\\ \hline
\beta & \beta & \alpha & \beta &  \beta & \alpha & \beta & \beta \\ 
\beta & \beta & \beta & \alpha &  \alpha & \beta & \beta & \beta \\ 
\alpha & \beta & \beta & \beta &  \beta & \beta & \beta & \alpha \\ 
\beta & \alpha & \beta & \beta &  \beta & \beta & \alpha & \beta \\ 
\end{matrix}
\label{eq:abl}
\end{equation}
where  the real  numbers $\alpha,  \beta \ge  0$ and  by normalization
$\alpha+3\beta=1$.   The   rows  correspond  sequentially   to  inputs
$ab=00,01,10,11$.   Here  $\alpha$  must  be less  than  the  unbiased
probability of $\frac{1}{4}$  to suppress the input  for which $E(a,b)
\ne (-1)^{a\overline{b}}$  in the contribution to  the CHSH inequality
(\ref{eq:bi}).

Letting  each $ab$  and each  $\lambda=\textbf{d}^{j_0}$ be  uniformly
probable, by Bayesian arguments we have from Eq. (\ref{eq:abl})
\begin{equation}
\rho(\lambda|ab)=\begin{matrix}
\textbf{d}^{0_0}  &  \textbf{d}^{1_0} &  \textbf{d}^{2_0}  &  \textbf{d}^{3_0}  &
\textbf{d}^{4_0} & \textbf{d}^{5_0} & \textbf{d}^{6_0} & \textbf{d}^{7_0}
\\ \hline
\betaf & \betaf & \alfaf & \betaf &  \betaf & \alfaf & \betaf & \betaf \\ 
\betaf & \betaf & \betaf & \alfaf &  \alfaf & \betaf & \betaf & \betaf \\ 
\alfaf & \betaf & \betaf & \betaf &  \betaf & \betaf & \betaf & \alfaf \\ 
\betaf & \alfaf & \betaf & \betaf &  \betaf & \betaf & \alfaf & \betaf \\ 
\end{matrix}.
\label{eq:lab}
\end{equation}
Applying this data to the discrete version of definition
(\ref{defn:Hal10}) we find
\begin{eqnarray}
F =1-|\alpha-\beta|=\frac{2+4\alpha}{3}.
\label{eq:F}
\end{eqnarray}
Further, from Table \ref{tab:bm0} and Eq. (\ref{eq:lab}), we find
\begin{equation}
P_{xy|ab} = \left\{\begin{array}{l||l|l|l|l}
ab=00 & 3\beta/2 & \alpha/2 & \alpha/2 & 3\beta/2 \\ \hline
ab=01 & 3\beta/2 & \alpha/2 & \alpha/2 & 3\beta/2 \\ \hline
ab=10 & \alpha/2 & 3\beta/2 & 3\beta/2 & \alpha/2 \\ \hline
ab=11 & 3\beta/2 & \alpha/2 & \alpha/2 & 3\beta/2,\\ 
\end{array}\right.
\label{eq:hallnosig}
\end{equation}
where  each  row represents  a  single  input  $ab$, and  the  columns
represent the outputs 00, 01, 10  and 11. We denote the correlation in
Eq.  (\ref{eq:hallnosig}) by $\textbf{P}^\ast_\mathcal{L}$.  For this,
we  find   that  for   each  input  $E(a,b)=   (-1)^{a\overline{b}}  (
3\beta-\alpha)=(-1)^{a\overline{b}} (1-2\alpha)$, so that
\begin{equation}
\Lambda = 4(1-2\alpha)
\label{eq:L}
\end{equation}
From Eqs.    (\ref{eq:F}) and (\ref{eq:L})    it   follows   that 
\begin{equation}
\Lambda = 2(4-3F).
\end{equation}
We note  that $\Lambda(F :=  1)=2$ and $\Lambda(F  := \frac{2}{3})=4$.
The quantum Cirelson bound of $2\sqrt{2}$ is reached when free will is
reduced to just $F = \frac{4-\sqrt{2}}{3} \approx 86\%$.

These agree with the results of \cite{Hal10}, but it may be noted that
we use a different set of boxes. Moreover, we do not require different
sets of  local boxes for reaching  the Cirelson bound or  reaching the
algebraic bound.

$\textbf{P}^\ast_\mathcal{L}$  resulting  is non-signaling.   This  is
because all boxes in (\ref{eq:abl})  are mixed with equal probability,
so that all inputs occur  with equal probability $(6\beta + 2\alpha)/8
=  \frac{1}{4}$, there  is no  correlation between  Alice's and  Bob's
inputs.     Substituting    the     $P_{xy|ab}$    data    from    Eq.
(\ref{eq:hallnosig}) into the  no-signaling conditions (\ref{eq:s()}),
one  finds  that   the  correlation  $\textbf{P}_\mathcal{L}^\ast$  is
non-signaling in that $S=0$.   

Consider  the operational  correlation $\textbf{P}^\circ_\mathcal{L}$,
formed  by   uniformly  mixing   only  the   boxes  $\textbf{d}^{0_0},
\textbf{d}^{1_0}, \textbf{d}^{2_0}$  and $\textbf{d}^{3_0}$.  Then
$\textbf{P}_\mathcal{L}^\circ$   will    satisfy   $P_{A|B}=P_A$   and
$P_{B|A}=P_B$ but fail the no-signaling conditions (\ref{eq:opnosig}).
In this  case $\rho(ab|\lambda)=\rho(\lambda|ab)$,  which is  given by
Eq.    (\ref{eq:abl}).     Thus,   from   Table    \ref{tab:bm0}   and
Eq. (\ref{eq:abl}), we find
\begin{equation}
P_{xy|ab} = \left\{\begin{array}{l||l|l|l|l}
ab=00 & 2\beta & \alpha & 0 & \beta \\ \hline
ab=01 & 3\beta & 0 & \alpha & 0 \\ \hline
ab=10 & \alpha & 2\beta & \beta & 0 \\ \hline
ab=11 & 3\beta & 0 & \alpha & 0,\\ 
\end{array}\right.
\label{eq:hallnosig0}
\end{equation}
which is  readily seen to  be signaling, with  $S=\beta-\alpha$, using
(\ref{eq:hallnosig0})   in  Eq.    (\ref{eq:s()}).   A   more  general
framework for free will  reduction, including deterministic boxes with
$\Lambda=+4$  given  in Table  \ref{tab:bm1},  is  discussed later  in
Section \ref{sec:srf}.

Ref.  \cite{CR11}  proposes  to identify  free  will  with  the
  requirement:
\begin{eqnarray}
P_{A|BY\lambda} &=& P_A,\nonumber\\
 P_{B|AX\lambda} &=& P_B,
\label{defn:AS}
\end{eqnarray}
where $\lambda$ has been substituted in place of ``static'' variables,
input SV $C$ and output SV  $Z$).  This would in effect generalize Eq.
(\ref{defn:Hal10}) by allowing for loss  of free will through explicit
dependence of Alice's input on Bob's input and vice versa.

Now,  the   definition  of   free  will  (\ref{defn:AS})   yields  the
ontological no-signaling conditions.  By Bayesian arguments:
\begin{equation}
P_{BX|A\lambda}   =
P_{B|AX\lambda}P_{X|A\lambda}   =   P_BP_{X|A\lambda}.
\label{eq:free1}
\end{equation}
and  again   
\begin{equation}
P_{BX|A\lambda}   =
P_{X|AB\lambda}P_{B|A\lambda} =  P_BP_{X|AB\lambda}.
\label{eq:free2}
\end{equation}
Equating the  r.h.s of Eqs. (\ref{eq:free1})  and (\ref{eq:free2}), we
derive  (\ref{eq:ontnosig}).  

Let $\mathcal{T}^+$ (resp., $\mathcal{T}^-$)  denote the causal future
(resp., causal past) in SR with  respect to some event \textbf{e}, and
$\overline{\mathcal{T}^+}$  (resp.,  $\overline{\mathcal{T}^-}$),  the
spacetime  region  outside $\mathcal{T}^+$  (resp.,  $\mathcal{T}^-$).
Further, $\mathcal{T}^0$ refers to  the ``twilight zone'' outside both
the  causal  future  and  past,  i.e., the  set  of  events  spacelike
separated   from    \textbf{e}.    Now,   if   in    accordance   with
\cite{CK06,*CK09,CR11},   the  scope   of   the  past   in  the   definition
(\ref{defn:AS}) to which  the conditioning SV's pertain  (e.g., $B$ or
$Y$ in  $P_{A|BY}$), is  taken to be  $\overline{\mathcal{T}^+}$, then
(as will be clarified later) this will prohibit certain ``beneficial''
ontological signaling. To avoid this dead-end, there are two responses
to this situation.

The first response  is that we may propose a  new covariant concept of
freedom which would only lead to the \textit{operational} no-signaling
conditions,  but   not  prohibit   ontological  signaling.    Such  an
``operational free will'', which  we call \textit{spontaneity}, is the
requirement  that Alice's  choice is  independent of  Bob's input  and
output, and vice  versa.  Thus, Alice's and  Bob's measurement choices
are spontaneous if:
\begin{eqnarray}
P_{A|BY} &=& P_A.\nonumber\\
P_{B|AX}&=&P_B,
\label{defn:spontan}
\end{eqnarray}
where the  scope of the  past is given  by $\overline{\mathcal{T}^+}$,
the  same   as  that  for  the   operational  no-signaling  conditions
(\ref{eq:opnosig}).      These    conditions     are    implied     by
(\ref{defn:spontan}), as seen by equating the rhs of
\begin{equation}
P_{BX|A} = P_{B|AX}P_{X|A}  = P_BP_{X|A},
\end{equation} 
and that of
\begin{equation}
P_{BX|A} = P_{X|AB}P_{B|A}  = P_BP_{X|AB}.
\end{equation}  
which  yields (\ref{eq:opnosig}).

The second response, to be studied in detail later below, is to retain
the definition  (\ref{defn:AS}), but  alter the scope  of the  past to
ensure  that useful  superluminal ontological  signaling is  not ruled
out. Thus, the scope of the past for (\ref{defn:AS}), and consequently
free will,  will not be  covariant.  For ontological  properties, this
does not matter.  What is required is a consistent and philosophically
coherent  definition of  free will  that conduces  to reproducing  the
operational  theory.  This  idea  will be  explicitly demonstrated  by
constructing an extension later below.

Note that as we have defined and ``scoped'' free will and spontaneity,
in a world which is non-signaling at the operational level, the former
implies the latter, but the converse  is not true.  By virtue of being
operationally     signaling,    \textbf{P}     described    by     Eq.
(\ref{eq:hallnosig0}), unlike that  described by (\ref{eq:hallnosig}),
stands in violation of spontaneity.  In a non-signaling world, loss of
spontaneity  in  choosing  inputs  can  only  come  through  a  signal
originating in the  past light cone, which would also  make the choice
unfree.

(But if the  world were such as to permit  superluminal signals at the
operational level, then one could violate (\ref{defn:spontan}) through
a $\mathcal{T}^0$ event lying in the  future as seen in some preferred
intertial reference frame.  In this case, we would have free will, but
not  spontaneity.   However,  this  pathological  situation  does  not
matter, since  in such  a world,  covariant concepts  like spontaneity
would be irrelevant.)

Therefore,   the   two  concepts   of   freedom,   namely  free   will
(\ref{defn:AS})  and spontaneity  (\ref{defn:spontan}), differ  in two
ways.  One is in the set of factors from which to be free, as a result
of  which free  will is  an  ontological concept,  but spontaneity  is
operational.  The second  way is in the scope of  the \textit{past} in
which  the  conditioning  SVs  are  located,  whereby  spontaneity  is
covariantly defined, whereas free will is not.

One point  worth noting with  regard to freedom, both  ontological and
operational,  is  that  whereas   the  freedom  conditions  imply  the
corresponding  no-signaling, the  converse  is not  true (see  below).
Thus,  a  `telepathic  signal'  can  be  generated  simply  through  a
correlation between Alice's and Bob's inputs, even when a conventional
operational  signal through  a correlation  between Alice's  input and
Bob's output or vice versa, is absent.

For    instance,   a    correlation   between    $A$   and    $B$   in
(\ref{defn:spontan}) but  none between  $A$ and $Y$  will not  lead to
signaling in  the sense of (\ref{eq:opnosig}),  but nevertheless leads
to a potential  communication (e.g., Alice finds that  she is inclined
to one or other input depending on Bob's remote choice).

To  illustrate this,  fix the  state to  be $\lambda=\textbf{d}^{4_0}$
defined in Table \ref{tab:bm0}, with the choice of inputs according to
scheme (\ref{eq:abl}).  For this data  we find the joint probabilities
$P_{AB=00}=P_{AB=10}  =P_{AB=11}=\beta$  and $P_{AB=01}=\alpha$.   The
marginal   probabilities  are   $P_{A=0}=P_{B=1}=\alpha+  \beta$   and
$P_{A=1}=P_{B=0}=2\beta$.   By   Bayesian  reasoning,  we   find  that
$P_{B=0|A=0}=\frac{P_{AB=00}}{P_{A=0}} = \frac{\beta}{\alpha +\beta}$,
which  equals $P_{B=0}$  if  and  only if  $\alpha=\beta=\frac{1}{4}$.
This dependence  of $P_B$ on input  $A$ entails that Alice  lacks free
will and spontaneity.  Suppose we take the operational state itself to
be $\textbf{d}^{4_0}$. Then $S=0$ and  yet Alice receives a telepathic
signal whereby she  discerns Bob's input by  examining her inclination
to choose one or the other input.

\section{Simulating singlet statistics \label{sec:simu}}

Suppose Alice  and Bob  measure input observables  labelled $a,  b \in
\{0,1\}$, respectively, on  a quantum state, and obtain  outputs $x, y
\in \{0,1\}$.  The general  2-input, 2-output correlation, represented
by  the  probability  vector  $\textbf{P} \equiv  P_{xy|ab}$,  can  be
decomposed into deterministic correlations,  which are elements of the
\textit{signaling   polytope}  $\mathcal{S}$   \cite{ASIQSA}.   Vector
$\textbf{P}$   has   16   entries,   governed   by   4   normalization
conditions. Thus the dimension of  $\mathcal{S}$ is 12. There are $4^4
= 256$  deterministic correlations  $\textbf{P}$, which  correspond to
the  extreme   points  of   $\mathcal{S}$.   Of  these,   sixteen  are
local-deterministic correlations, and  the remaining 240 deterministic
correlations are  not local.  The no-signaling  polytope $\mathcal{N}$
\cite{BLM+05} is an 8-dimensional  polytope within $\mathcal{S}$, with
vertices  given by  the  16 local-deterministic  correlations and  the
eight PR boxes,  which violate CHSH inequalities  \cite{CHSH} to their
algebraic maximum of 4.

\subsection{A polytope fragment $\mathfrak{F}^P$}

For our purpose, we do not  need to consider all of $\mathcal{S}$, but
the fragment of it, which  we denote $\mathfrak{F}^P$, obtained as the
convex  hull   of  eight  0-bit  boxes   $\textbf{d}^{j_0}$  in  Table
\ref{tab:bm0},  for  which $\Lambda=+2$,  and  the  eight 1-bit  boxes
$\textbf{d}^{j_1}$ in Table \ref{tab:bm1}, for which $\Lambda=+4$.  We
shall refer to any $\textbf{P}$  in $\mathfrak{F}^P$ as a ``$C$-box''.
These  boxes,   given  in   Tables  \ref{tab:bm0}   and  \ref{tab:bm1}
respectively, constitute the extreme  points of $\mathfrak{F}^P$.  Our
study  below  can   be  easily  extended  to  a   larger  fragment  of
$\mathcal{S}$,  but  $\mathfrak{F}^P$  is sufficient  in  the  present
context.  Moreover, any \textbf{P} in  $\mathfrak{F}^P$ can be used as
a resource  to simulate the statistics  of a singlet.

Any  $\textbf{P} \in  \mathfrak{F}^P$, not  necessarily non-signaling,
can be decomposed as:
\begin{equation}
P_{xy|ab}  =  \sum_{j=0}^7   p^{0}_j  d^{j_0}_{xy|ab}  +  \sum_{j=0}^7
p^{1}_j d^{j_1}_{xy|ab}
\label{eq:HV}
\end{equation}
where $p^k_j\ge0$.  Let $p_0\equiv\sum_j p^{0}_j$ and $p_1\equiv\sum_j
p^{1}_j$.    Normalization  requires   $p_0+p_1=   1$.   The   optimal
decomposition   for  $P_{xy|ab}$   is  one   that  minimizes   in  Eq.
(\ref{eq:HV})  the quantity  $p_1$, which,  as we  show below,  is the
average  communication   cost  $C$  for  simulating   the  correlation
\textbf{P}.

Decomposition       (\ref{eq:HV})       defines       a       protocol
$\mathfrak{S}(\chi,\mathcal{R})$    to    simulate    \textbf{P}    in
$\mathfrak{F}^P$. Let  $\chi$ represent pre-shared  randomness between
two  simulating   parties  (designated  ``Alice''  and   ``Bob'')  and
$\mathcal{R}$, a  communicated resource  that depends on  Alice's free
choice of  $a$ and her  outcome $x$  \cite{PKP+10}.  In this  work, we
take  her  outcome  information  to be  restricted  to  $\chi$,  while
information about her input will be restricted to $\mathcal{R}$.

The execution of $\mathfrak{S}(\chi,\mathcal{R})$ proceeds as follows:
Alice   and    Bob   pre-share    a   4-bit   stream    $\chi   \equiv
\chi_0\chi_1\cdots\chi_k\cdots$, where  each 4-bit specifies  which of
the  16 strategies  $\textbf{d}^{j_0}$ or  $\textbf{d}^{j_1}$ will  be
used.  The  fraction $p_0$  of zero-bit  strategies and  the remaining
fraction $p_1$  of 1-bit strategies  will be pre-decided  according to
the  level  of  inequality  violation  sought.   When  the  $k$th  run
corresponds to  a zero-bit strategy $\textbf{d}^{j_0}$,  Alice and Bob
freely (i.e., independently of the $\chi_k$) choose inputs $a$ and $b$
respectively,  and  read-out outputs  $x$  and  $y$ according  to  the
pre-shared $\textbf{d}^{j_0}$.   When the  run corresponds to  a 1-bit
strategy $\textbf{d}^{j_1}$, again both freely choose their respective
input.    Alice    outputs   $x$    according   to    the   pre-shared
$\textbf{d}^{j_1}$,  and further  she  transmits to  Bob the  resource
$\mathcal{R}$, which  in this case  is the 1-bit information  $a$. Bob
computes  $y$ that  would return  $\Lambda=+4$  given $b,  a$ and  the
pre-shared strategy $\textbf{d}^{j_1}$ for that run, i.e., he computes
$y =  a\cdot\overline{b} \oplus  x$.  Clearly,  this protocol  gives a
practical realization of a  decomposition (\ref{eq:HV}) which produces
on  average  $\Lambda =  4p_1  +  2(1-p_1)=2(1+p_1)$. Note  that  this
reaches the algebraic maximum of $\Lambda=+4$ at $p_1=1$ and the local
maximum $\Lambda=+2$ at $p_1=0$.

\subsection{Optimal protocol\label{sec:protocol}}

The operational no-signalling  conditions for the two-input-two-output
situation are given by:
\begin{subequations}
\begin{eqnarray}
P_{0y|00}  +  P_{1y|00}  &=&  P_{0y|10}  + P_{1y|10}  \\  P_{0y|01}  +
P_{1y|01} &=& P_{0y|11} +  P_{1y|11} \label{eq:nosig12} \\ P_{x0|10} +
P_{x1|10} &=& P_{x0|11} +  P_{x1|11} \label{eq:nosig47} \\ P_{x0|00} +
P_{x1|00} &=& P_{x0|01} + P_{x1|01} \label{eq:nosig56},
\end{eqnarray}
\label{eq:sigmno}
\end{subequations}
Allowing  for the  general violation  of no-signaling,  the conditions
(\ref{eq:sigmno}) become:
\begin{subequations}
\begin{eqnarray}
P_{00|00}    +    P_{10|00}   &=&    P_{00|10}    +   P_{10|10}  - 
\delta_{I} \nonumber\\
P_{01|00}    +    P_{11|00}   &=&    P_{01|10}    +   P_{11|10} +
\delta_{I} 
\label{eq:sig03} 
\\ 
P_{00|01}  + P_{10|01} &=& P_{00|11} +
P_{10|11}  + \delta_{II} \nonumber\\
P_{01|01}  + P_{11|01} &=& P_{01|11} +
P_{11|11}  - \delta_{II} \label{eq:sig12}  
\\ 
P_{00|10}  + P_{01|10}
&=&   P_{00|11}   +   P_{01|11}   +   \delta_{III} \nonumber \\
P_{10|10}  + P_{11|10} 
&=&   P_{10|11}   +   P_{11|11} -   \delta_{III} 
  \label{eq:sig47}
\\  P_{00|00}  +  P_{01|00}  &=&  P_{00|01}  +  P_{01|01}  +
\delta_{IV} \nonumber\\
P_{10|00}  +  P_{11|00}  &=&  P_{10|01}  +  P_{11|01}  -
\delta_{IV} \label{eq:sig56},
\end{eqnarray}
\label{eq:sigmn}
\end{subequations}
where $\delta_{j}$'s ($j  \in \{I, II, III,  IV\}$) quantify violation
of   the   no-signaling   condition.    Eqs.    (\ref{eq:sig03})   and
(\ref{eq:sig12}) indicate  signaling from  Alice to Bob,  whereas Eqs.
(\ref{eq:sig47}) and (\ref{eq:sig56}) indicate signaling from Alice to
Bob.  Further, we have:
\begin{equation}
S(\textbf{P})  \equiv \max_{j} |\delta_j| 
\label{eq:delta}
\end{equation}
from Eq. (\ref{eq:s()}).

For   the   fragment    $\mathfrak{F}^{P}$   of   two-input-two-output
correlations, one finds using Table \ref{tab:bm1} that:
\begin{eqnarray}
\delta_{I} &\equiv& p^1_3 - p^1_0\nonumber\\
\delta_{II} &\equiv& p^1_2 - p^1_1\nonumber\\  
\delta_{III}  &\equiv&  p^1_7 -  p^1_4\nonumber\\  
\delta_{IV} &\equiv& p^1_6 - p^1_5. 
\label{eq:deltaj}
\end{eqnarray}
Each non-vanishing $\delta_j$ can thus  be interpreted as an imbalance
in  the probability  with  which  a box-antibox  pair  of 1-bit  boxes
appears   in   decomposition   (\ref{eq:HV}).   If   for   some   $j$,
$\delta_j\ne0$,  then operational  no-signaling (\ref{eq:opnosig})  is
violated.

For a  general (possibly signaling) $\textbf{P}  \in \mathfrak{F}^P$ ,
we   now   show   how  to   construct   decomposition   (\ref{eq:HV}).
Eq. (\ref{eq:bi}) may be expanded as:
\begin{eqnarray}
\Lambda(\textbf{P}) &=& (P(a=b|00) + P(a= b|01) + P(a \ne b|10)  \nonumber \\
&+& P(a=b|11))
-  (P(a\ne b|00) + P(a\ne b|01) \nonumber \\ &+& P(a= b|10) 
+ P(a\ne b|11)).
\label{eq:lg+}
\end{eqnarray} 
The contribution of  the negative signs for  $\Lambda(\textbf {P})$ in
Eq. (\ref{eq:lg+}) is only from the $\textbf{d}^{j_0}$ boxes, and
fixes the eight $p^0_j$'s, as follows:
\begin{eqnarray}
p^0_0 = P_{00|10} &;&~ p^7_0 = P_{11|10} \nonumber \\
p^1_0 = P_{10|11} &;&~ p^6_0 = P_{01|11} \nonumber \\
p^2_0 = P_{01|00} &;&~ p^5_0 = P_{10|00} \nonumber \\
p^3_0 = P_{10|01} &;&~ p^4_0 = P_{01|01}.
\end{eqnarray}
The positive terms  are constructed with both  $\textbf{d}^{j_0} $ and
$\textbf{d}^{j_1}$  deterministic  boxes.    For  example,  using
Tables \ref{tab:bm0} and  \ref{tab:bm1}, $P_{00|00} = p^0_0  + p^0_1 +
p^0_4 + p^1_0 + p^1_2 + p^1_4 + p^1_6$.

Substituting  for  the $p^0_j$  terms  as  above  gives the  r.h.s  of
(\ref{eq:new1})  below, which,  with the  normalization and  signaling
conditions (\ref{eq:sigmn}), gives the r.h.s in (\ref{eq:new2}):
\begin{subequations}
\begin{eqnarray}
\sum_{j  = 0,2,4,6}  p^1_j  &=&  P_{00|00}  - P_{00|10}  - P_{10|11}  -
P_{01|01} \label{eq:new1} \\
&=& \frac{1}{2}(C_\Lambda  - \delta_{I}  +
\delta_{II}     -     \delta_{III}     +    \delta_{IV}),
\label{eq:new2}
\end{eqnarray}
\label{eq:new}
\end{subequations}
where  $C_\Lambda \equiv  \frac{\Lambda}{2}-1$.  Substituting  for the
$\delta_j$'s in  Eq.  (\ref{eq:new})  using Eq.  (\ref{eq:deltaj}), we
find that
\begin{equation}
p_1=\sum_{j=0}^7  p^1_j =  C_\Lambda.
\label{eq:new0}
\end{equation}
Eq.     (\ref{eq:new0})   together    with    the   four    conditions
(\ref{eq:deltaj}) constitute five independent constraints on the eight
$p^1_j$'s, leaving three  free terms $p^1_j$ as  expected, since there
are 15 probabilities $p^k_j$ in (\ref{eq:HV}), and $\mathfrak{F}^P$ is
12-dimensional. In  the non-signaling  case, we set  all $\delta_j=0$,
and  find   that  all  box-antibox  $\textbf{d}^j_1$   boxes  must  be
balanced.  Our  protocol $\mathfrak{S}(\chi,\mathcal{R})$  generalizes
the protocol for non-signaling \textbf{P} given in Ref. \cite{Pir03}.

We now show that any  decomposition (\ref{eq:HV}) as determined by the
method      above     gives      an      optimal     protocol      for
$\textbf{P}\in\mathfrak{F}^P$.   Consider an  arbitrary \textbf{P}  in
$\mathcal{S}$  that can  be  decomposed  in terms  of  0-bit or  1-bit
deterministic  boxes.  For  each  of these  boxes, the  only
possible values  of $\Lambda$ in  Eq. (\ref{eq:bi}) are $\pm4,  \pm 2,
0$,  with $\Lambda=\pm4,0$  (resp.,  $\Lambda=\pm2$) corresponding  to
1-bit (resp.,  0-bit) boxes.   Let the  corresponding probability
with which they  appear in a general  decomposition like (\ref{eq:HV})
be denoted $q_{\pm4}, q_{\pm2}$ and $q_0$. Now
\begin{eqnarray}
C_\Lambda &=& \frac{4q_{+4} + 2q_{+2} - 2q_{-2} - 4q_{-4}}{2}-1 \nonumber\\
   &=& 2\left(q_{+4}-q_{-4}\right) + \left(q_{+2}-q_{-2}\right) - 1 \nonumber \\
   &\le& 2\left(q_{+4}+q_{-4}\right) + \left(q_{+2}+q_{-2}\right) - 1 \nonumber\\
      &=& q_{+4}+q_{-4} -q_0 \nonumber \\ \label{eq:ineq4}
      &\le& q_{+4}+q_{-4} + q_0 = p_1,
\label{eq:ineq}
\end{eqnarray}
where  $p_1$ is  the  average  number bits  required  to simulate  the
protocol.   Since  $C$  is  $p_1$ minimized  over  all  decompositions
(\ref{eq:HV}),  Eq.   (\ref{eq:ineq})  implies $C\ge  C_\Lambda$.   As
$\mathfrak{S}(\chi,\mathcal{R})$ is implemented  with $C_\Lambda$ bits
of  average communication  in view  of Eq.  (\ref{eq:new0}), and  thus
attains this lower bound on $C$, the protocol is optimal.

In fact, any protocol associated with decomposition (\ref{eq:HV}) will
be optimal for $\mathfrak{F}^P$,  since this construction ensures that
$q_{-4}=q_0=0$, and  therefore that all  1-bit boxes used  in the
simulation contribute  maximally (with $\Lambda=+4$) to  the violation
of (\ref{eq:bi}).

\subsection{From $\mathfrak{F}^P$
to singlet statistics\label{sec:simux}}

Although   the  C-box   defined  above   is  a   rather  simple
  two-input-two-output correlation,  it suffices as the  resource that
  Alice  needs to  communicate to  Bob  in order  to simulate  singlet
  statistics.   This  task   requires  that,   given  random   vectors
  $\hat{n}_A$  and $\hat{n}_B$,  respectively, Alice  and Bob  produce
  outputs $\pm  1$, such that  the product average  equals $-\hat{n}_A
  \cdot \hat{n}_B$.   If we  relabel the respective  outcomes, denoted
  $\textbf{n}_A$ and  $\textbf{n}_B$, to take  on values 0 or  1, then
  the simulation must reproduce:
\begin{equation}
 \left\langle\textbf{n}_A \oplus \textbf{n}_B\right\rangle  = \frac{1}{2}(1  +
   \hat{n}_A \cdot \hat{n}_B),
\label{eq:syngcor}
\end{equation}
where the expectation  value is denoted by  the angle brackets.

We  now  briefly  recapitulate  from  \cite{ASQIC}  how  the  $C$-box,
supplemented with  other pre-shared randomness, denoted  $\chi^+$, can
be  used as  a ``sub-routine''  to simulate  singlet statistics.   The
randomness  $\chi^+ \equiv  \{\eta_1, \eta_2\}$,  where each  $\eta_j$
($j=1, 2$)  is a uniformly distributed,  independent direction vector.
Given   the  two   arbitrary  angles   $\hat{n}_A$  and   $\hat{n}_B$,
respectively, Alice  computes $\upsilon_A  = \mbox{sgn}\left(\hat{n}_A
\cdot    \hat{\eta}_1\right)    \oplus   \mbox{sgn}(\hat{n}_A    \cdot
\hat{\eta}_2)$, which she inputs into the resource $C$-box with $C=1$.
We use the notation  that $$\textrm{sgn}(m) = \left\{\begin{array}{cc}
0 & \rightarrow m < 0\\ 1 & \rightarrow m \ge0
\end{array}.\right.$$
She obtains output $\alpha$ from this resource, from which she derives:
\begin{equation}
\textbf{n}_A = \alpha\oplus \mbox{sgn}(\hat{n}_A\cdot\hat{\eta}_1).
 \end{equation}
Bob computes $ \upsilon_B = \mbox{sgn}(\hat{n}_B\cdot\hat{\eta}_{(+)})
\oplus        \mbox{sgn}(\hat{n}_B\cdot\hat{\eta}_{(-)})$,       where
$\hat{\eta}_{(\pm)}=\hat{\eta}_1\pm      \hat{\eta}_2$.      Inputting
$\upsilon_B$  into the  $C$-box  he received  from  Alice, he  obtains
outcome $\beta$, from which he computes:
\begin{equation}
\textbf{n}_B = \beta\oplus \mbox{sgn}(\hat{n}_B\cdot\hat{\eta}_{(+)})\oplus1.
 \end{equation}
By direct substitution, this yields
\begin{eqnarray}
\textbf{n}_A \oplus   \textbf{n}_B    &   =    &   \alpha   \oplus    \beta   \oplus
\mbox{sgn}(\hat{n}_A\cdot\hat{\eta}_1)                          \oplus
\mbox{sgn}(\hat{n}_B\cdot\hat{\eta}_{(+)})\oplus1  \nonumber \\
&=&  \xi\upsilon
\oplus          \mbox{sgn}(\hat{n}_A\cdot\hat{\eta}_1)         \oplus
\mbox{sgn}(\hat{n}_B\cdot\hat{\eta}_{(+)})\oplus1.
 \end{eqnarray} 
It can be shown that the above correlation can be used Alice and
Bob 
to reproduce singlet  correlations (\ref{eq:syngcor}) employing the
method described in Ref. \cite{CGM+05}.

\subsection{Complementarity in resources for simulating
singlet statistics \label{sec:cmp}}

A  complementarity  is known  to  exist  between signaling  and  local
randomness in  the resources required  to be communicated in  order to
simulate a ``$C$-box'', of the form:
\begin{equation}
S_R + 2I_R \ge C_\Lambda.
\label{eq:s2ic}
\end{equation}
The  proof (which  appears in  detail in  \cite{ASQIC}) is  briefly as
follows.   It   can  be  shown  that   $S  +  2I  \ge   p_1$  for  any
$\textbf{P}\in\mathfrak{F}^P$.  In Section  \ref{sec:protocol}, we saw
that  any  $C$-box,   by  virtue  of  optimality,   satisfies  $p_1  =
C_\Lambda$.  Ineq.  (\ref{eq:s2ic}) then follows.

Setting $C_\Lambda$ to  the maximal value of 1, gives  the $C$-box for
which
\begin{equation}
S_R + 2I_R \ge 1.
\label{eq:senior}
\end{equation}
In Section \ref{sec:simux}, we showed that this maximal $C$-box can be
used as  the communicated  resource that  suffice to  simulate singlet
statistics, with supplementary pre-shared  information $\chi^+$ in the
form  of unbiased  bits.   We shall  denote  this extended  simulation
protocol also  by $\mathfrak{S}(\chi,\chi^+,\mathcal{R})$.   Where $S$
and $I$  refer to  a $C$-box  used as a  resource to  simulate singlet
statistics, for  clarity, we shall  subscript them with an  $R$, i.e.,
refer to them as $S_R$ and $I_R$.

This protocol can  also be shown to be optimal  for simulating singlet
statistics in the sense  of minimizing communicated bits \cite{BKP06}.
Accordingly,  Ineq.   (\ref{eq:senior})  can   be  considered  as  the
complementarity of communicated resources required to simulate singlet
statistics.   The  case  $(S_R=1,  I_R=0)$ corresponds  to  the  1-bit
Toner-Bacon protocol \cite{TB03} for this  task, and the case $(S_R=0,
I_R=\frac{1}{2})$ corresponds to a PR-box based protocol \cite{CGM+05}
for the same task.

\subsection{$\textbf{P}^\ast_\mathcal{L} \in \mathfrak{F}^P$
\label{sec:redu}}

We note that $\rho(ab|\lambda)$ in Eq. (\ref{eq:abl}) can be expressed
as the sum
\begin{equation}
\rho(ab|\lambda) = 4\alpha\cdot\rho_0(ab|\lambda) + 
(1-4\alpha)\cdot\rho^\ast(ab|\lambda),
\end{equation} 
where $\rho_0(ab|\lambda) := \frac{1}{4}$ for all inputs $ab$ and
all $\lambda = \textbf{d}^{j_0}$, and
\begin{equation}
\rho^\ast(ab|\lambda)=\begin{matrix}
\textbf{d}^{0_0}  &  \textbf{d}^{1_0} &  \textbf{d}^{2_0}  &  \textbf{d}^{3_0}  &
\textbf{d}^{4_0} & \textbf{d}^{5_0} & \textbf{d}^{6_0} & \textbf{d}^{7_0}
\\ \hline
\delta & \delta & 0 & \delta &  \delta & 0 & \delta & \delta \\ 
\delta & \delta & \delta & 0 &  0 & \delta & \delta & \delta \\ 
0 & \delta & \delta & \delta &  \delta & \delta & \delta & 0 \\ 
\delta & 0 & \delta & \delta &  \delta & \delta & 0 & \delta \\ 
\end{matrix}
\label{eq:abl+}
\end{equation}
where $\delta\equiv(\beta-\alpha)/(1-4\alpha)=\frac{1}{3}$.  Under the
uniform mixing  of $\rho_0(ab|\lambda)$ over all  $\lambda$ yields the
local distribution $\textbf{P}_\mathcal{L}$.  On the other hand, under
uniform   mixing  of   $\rho^\ast(ab|\lambda)$  over   all  $\lambda$,
referring to (\ref{eq:hallnosig}), a PR box in $\mathfrak{F}^P$, which
can be considered the equal mixture of  any pair of box and antibox in
Table \ref{tab:bm1}.

Therefore  a  uniform mixture  of  the  $\textbf{d}^{j_0}$ boxes  with
reduced free  will mode according  to (\ref{eq:abl}) is  equivalent to
the  protocol in  the free  mode $\mathcal{F}$,  obtained in  the with
0-bit  boxes  mixed uniformly  with  total  probability $4\alpha$  and
combined with  a PR box  with probability weight  $(1-4\alpha)$.  Thus
$\textbf{P}^\ast_\mathcal{L}$ lies in $\mathfrak{F}^P$. 

\section{Complementarity incorporating free will \label{sec:srf}}   

To  incorporate free  will in  the  context of  above simulations  and
complementarity, we shall take $\lambda (\in \{\textbf{d}^{j_k}\})$ to
refer  to  simulation  strategies.   Correlations  between  strategies
$\lambda$   and   measurement   choices,   described   by   nontrivial
$\rho(\lambda|ab)$, will lead to a reduction in free will.  The method
described earlier in which only $\textbf{d}^{j_0}$ strategies are used
when reducing free will will be referred to as the $\mathcal{L}$ mode.
As one way to include $\textbf{d}^{j_1}$ strategies, we introduce mode
$\mathcal{F}$, which  uses only these  1-bit strategies.  Since  it is
already true for these strategies  that $\Lambda=+4$, they are applied
freely, requiring  no biasing of  input.  The mode that  combines both
$\mathcal{L}$ and $\mathcal{F}$ is denoted $\mathcal{LF}$, details for
which are discussed below.


In the $\mathcal{LF}$ mode, we fix  the probability of the 0-bit boxes
to be $l$, and  those of the 1-bit boxes to $1-l$.   Thus, in place of
Eq. (\ref{eq:lab}) we have for the 0-bit boxes:
\begin{equation}
\rho(\lambda|ab)=\begin{matrix}
\textbf{d}^{0_0}  &  \textbf{d}^{1_0} &  \textbf{d}^{2_0}  &  \textbf{d}^{3_0}  &
\textbf{d}^{4_0} & \textbf{d}^{5_0} & \textbf{d}^{6_0} & \textbf{d}^{7_0}
\\ \hline
\lf\beta & \lf\beta & \lf\alpha & \lf\beta &  \lf\beta & \lf\alpha & \lf\beta & \lf\beta \\ 
\lf\beta & \lf\beta & \lf\beta & \lf\alpha &  \lf\alpha & \lf\beta & \lf\beta & \lf\beta \\ 
\lf\alpha & \lf\beta & \lf\beta & \lf\beta &  \lf\beta & \lf\beta & \lf\beta & \lf\alpha \\ 
\lf\beta & \lf\alpha & \lf\beta & \lf\beta &  \lf\beta & \lf\beta & \lf\alpha & \lf\beta \\ 
\end{matrix}.
\label{eq:abl1}
\end{equation}
For  any $\lambda  := \textbf{d}^{j_1}$  boxes,  and any  of the  four
inputs, we have
\begin{equation}
\rho(\lambda|ab)=
\frac{1}{8}(1-l) 
\label{eq:abl2}
\end{equation}
Applying (\ref{eq:abl1}) and  (\ref{eq:abl2}) into (\ref{defn:Hal10}),
we find that
\begin{eqnarray}
F =1-l|\alpha-\beta|=1-\frac{l}{3}(1-4\alpha).
\label{eq:FB}
\end{eqnarray}
Further, we find $E(a,b) = (-1)^{a\overline{b}} (1 - 2\alpha l)$, so that 
\begin{equation}
\Lambda = 4(1-2\alpha l).
\label{eq:raf}
\end{equation}
This gives:
\begin{equation}
\Lambda = 4 - 6(F-1) - 2l,
\label{eq:freegen}
\end{equation}
using (\ref{eq:FB}) to replace $\alpha$ by $F$ in (\ref{eq:raf}).
For  the  general scenario  defined  above,  the
complementarity is altered, as discussed below.


\begin{thm}
To   simulate   a   non-signaling   two-input-two-output   correlation
\textbf{P},  the signaling  and local  randomness in  the communicated
resource $\mathcal{R}$ must satisfy:
\begin{equation}
S_R + 2I_R \ge C_\Lambda - 3(1-F)
\label{eq:main}
\end{equation}
with $1-\frac{C}{3} \le F \le 1$.
\label{thm:main}
\end{thm} 
\textbf{Proof.}    Let  the   fraction   of  the   local-deterministic
simulation  strategies (giving  $\Lambda=+2$) be  $l$.  Then  the CHSH
quantity   $\Lambda$   in  Eq.    (\ref{eq:bi})   is   given  by   Eq.
(\ref{eq:freegen}).
(When  only local-deterministic  strategies are  used ($l=1$)  we have
$\Lambda=2$ when $F=1$.  Reducing free will $F$ to $\frac{2}{3}$ leads
to the  algebraic maximum violation of  (\ref{eq:bi}) of $\Lambda=4$.)
From Eq.  (\ref{eq:freegen}), we have the CHSH inequality violation
\begin{eqnarray}
C_\Lambda  \equiv \frac{\Lambda}{2}-1 
= 4-3F-l,
\label{eq:uptoC}
\end{eqnarray} 
provided by  this mixture of reduction  of free will and  use of 1-bit
strategies.  From Eq. (\ref{eq:uptoC})
\begin{equation}
C_\Lambda = 3(1-F) + 1-l.
\label{eq:exCess}
\end{equation}
For  the  local-deterministic  part,  there  is  no  complementaristic
constraint  on signaling  and randomness.  However, for  the remaining
part, one requires the transmission of a bit with probability $1-l$.

Thus the  communicated  resource  for
simulation corresponds to a  resource $\mathcal{R}$ with communication
cost $C=1-l$.  Therefore by (\ref{eq:s2ic})
\begin{eqnarray}
S_R + 2I_R &\ge& 1-l \nonumber \\
  &=& C_\Lambda - 3(1-F),
\label{eq:asto}
\end{eqnarray}
using Eqs.   (\ref{eq:exCess}), which is Eq.  (\ref{eq:main}).  \hfill
$\blacksquare$
\bigskip

The interpretation  of Theorem \ref{thm:main} is  that increasing free
will imposes a larger demand on  the other two `nonlocal resources' of
$S_R$ and $I_R$.  From Eq.  (\ref{eq:main}),  it is seen that if $F=1$
then we  recover the complementarity, and  if free will is  lowered to
$1-\frac{C}{3}$ (or lower) then $S_R +  2I_R \ge 0$, i.e., there is no
bound  on   $S$  and  $I$.    Since  Bell's  theorem   corresponds  to
$S_R+2I_R>0$  \cite{ASQIC}, thus  Eq.  (\ref{eq:main})  represents the
strengthened form of Bell's theorem at a given level of free will.

Let $\textbf{P}^\ast_\mathcal{LF}$ define  the correlation obtained in
this scenario.  Because  it is constructed as a  convex combination of
$\textbf{P}^\ast_\mathcal{L}$   and   $\textbf{d}^{j_1}$   boxes,   by
convexity of $\mathfrak{F}^P$,  $\textbf{P}^\ast_\mathcal{LF}$ lies in
$\mathfrak{F}^P$.   Consequently, $\textbf{P}_\mathcal{LF}^\ast$  will
in general be a biased PR  box, which is characterized by $\Lambda=4$,
but  may   have  nonvanishing  signal  $S$.    When  $F:=\frac{2}{3}$,
$\textbf{P}_\mathcal{L}^\ast$ will be the PR box.

\section{Turning a simulation protocol into
an ontological extension\label{sec:srx}}

Intuitively, an ontological extension for  QM is like a ``simulation''
performed by  Nature.  Hence,  a natural question  is whether  one can
convert  a   simulation  protocol  such  as   $\mathfrak{S}$  into  an
ontological mechanism underlying quantum mechanics.  There seems to be
an  obvious  impediment militating  against  such  a conversion.   The
simulation  can  only reproduce  a  \textit{timeless}  version of  the
physical experiment, since a  timed version would require superluminal
communication  of   resource  $\mathcal{R}$   if  Alice's   and  Bob's
measurements are  spacelike separated.  Now if  $S_R=0$, this presents
no major difficulty.

However, if  the resource has reduced  randomness ($I_R<\frac{1}{2}$),
then for  maximum $F$, we  see from Eq. (\ref{eq:main})  that $S_R>0$.
Such a nonvanishing spacelike signal would not only pose a problem for
relativistic causality, but also for  the covariant definition of free
will  (\ref{defn:AS})  with  the  scope   of  the  past  taken  to  be
$\overline{\mathcal{T}^+}$.   This definition  would  entail that  the
no-signaling (\ref{eq:ontnosig})  should hold for  spacelike separated
events.

The concept of freedom in measurment settings in the simulation can be
readily  transferred  to  the  concept   of  free  will  in  practical
experiments by interpreting the simulation strategies $\lambda$ as the
ontological  state  in  the  ontic  support  of  a  given  operational
state.  We  will  thus  use  the   same  symbol  $F$  to  denote  free
will. However, there is the above issue of relativistic causality that
must be addressed, as we do below.

To try  to avoid this  difficulty with  defining free will,  one might
consider restricting  the scope  of the  past in  (\ref{defn:AS}) from
$\overline{\mathcal{T}^+}$ to $\mathcal{T}^-$,  the causal past.  This
would mean that  Alice and Bob can freely choose  their inputs, with a
superluminal signal connecting Alice's input  and Bob's outcome in the
extension.  But  this would allow  us to  choose a reference  frame in
which Alice's  measurement succeeds Bob's outcome,  whereby the signal
received by Bob becomes a restriction  on her free will.  Thus in this
case, free  will becomes an incoherent  concept.  These considerations
bring  out the  difficulty with  defining free  will covariantly  in a
nonlocal world,  and may at first  glance lead one to  conclude that a
predictively superior extension would contradict no-signaling and free
will.

Careful  inspection   shows  that  this  argument   assumes  that  the
ontological features  in an  extension of QM  should be  covariant and
subject to relativistic  causality.  But there is no  reason to assume
that the causal  structure of the spacetime in which  the extension is
set should  not be concommittantly  ``extended'' in some  way.  Worded
differently,  when the  protocol is  so converted,  we intend  for the
simulation resources  to carry  over to  the ontological  (rather than
operational) resources, according to Eq. (\ref{eq:conversion}).

Crucially,  $S_R$ does  not take  on significance  of the  operational
signal $S$,  which would be  prohibited by relativity.  And  as noted,
superluminal $S_\lambda$ is not prohibited by SR.  The concept of free
will  goes through  as it  is by  re-interpreting simulation  strategy
$\lambda$ as  HV $\lambda$.  We  now show how  protocol $\mathfrak{S}$
can be  embedded in such an  extended SR to construct  a non-covariant
extension of QM that reproduces a covariant operational theory.

An ``extension  of SR''  (or, SRX), briefly  mentioned earlier,  is an
ontological  model  of  events  that  presents  a  causal  account  or
``story''  of   nonlocal  quantum  correlations,   possibly  requiring
superluminal  signaling,  but   consistent  with  \textit{operational}
no-signaling.  The type of SRX that we are concerned with here are the
$v$-causal models \cite{SBB+08}.  More specifically, an SRX is a model
of events,  equipped with a causal  structure, in which the  ``cone of
causally connected events" is wider than that of the light cone of SR.

\subsection{Ontological extension for special relativity
\label{sec:srX}}

In an SRX,  at each event \textbf{e}, we define  ``the twilight zone''
$\mathcal{T}^0(\textbf{e})$  as   the  set  of  events   not  causally
connected to \textbf{e}.   The causally connected events  lying to the
future  (resp., past)  constitute  the  causal future  $\mathcal{T}^+$
(resp.,    causal     past    $\mathcal{T}^-$).      By    definition,
$\overline{\mathcal{T}^0}  (\textbf{e})   =  \mathcal{T}^+(\textbf{e})
\cup \mathcal{T}^-(\textbf{e})$.  SR is the trivial SRX, equipped with
the Minkowski causal structure, given by the usual light cones.

In a non-trivial SRX,  the twilight zone $\mathcal{T}^{0}(\textbf{e})$
at each event \textbf{e} is contained within the SR twilight zone:
\begin{equation}
\mathcal{T}^{0\prime}  (\textbf{e})  \subseteq \mathcal{T}^0_{\rm  SR}
(\textbf{e}),
\label{defn:srx}
\end{equation}
i.e., events not causally connected in SR may be causally connected in
an  SRX  (or, ``X-causally  connected''),  but  events not  X-causally
connected will not be causally connected  (in SR).  The trivial SRX is
SR, while the \textit{Newtonian} SRX is  one in which the spacetime is
Newtonian, and any pair of events is causally connected.


It will be convenient for us to consider a single-parameter continuous
family of  SRX's.  We designate  as the ``preferred  reference frame''
(PRF) a particular inertial reference frame.  We define $v_{\lambda}$,
or ``the speed  of the ontological signal'', as the  maximum extent of
spacetime through  which one SV  can causally influence another  in an
SRX.  Here $v_{\lambda}$  is  similar  to the  concept  of ``speed  of
quantum    information''     \cite{ZBT+01,Gar02,SBB+08}.     In    SR,
$v_{\lambda}=c$.  We  generate a  family of  SRX's by  continuouly and
symmetrically widening the causal (i.e., future or past) cones, making
$v_{\lambda}$ increase from  $c$ to $\infty$, as seen in  the PRF.  We
will refer to the  causal cones of the SRX as  ``X-cones''.  It can be
shown,  in view  of (\ref{defn:srx}),  that $v_{\lambda}\ge  c$ in  an
arbitrary SRX in this family.

Let $\theta_{\lambda}$ be the opening angle of the X-cone with respect
to   the  vertical,   as  seen   in  the   PRF.   SR   corresponds  to
$\theta_{\lambda}=\frac{\pi}{4}$, while for an  arbitrary SRX, we have
$\frac{\pi}{4}   \le   \theta_{\lambda}  \le   \frac{\pi}{2}$.    The
\textit{Newtonian} extension corresponds to  $v_{\lambda} = \infty$ and
$\theta_{\lambda}=\frac{\pi}{2}$.  The twilight zone of any SRX in this
family is denoted  $\mathcal{T}^0_{[\theta_{\lambda}]}$.  The Newtonian
SRX is characterized by a vanishing twilight zone:
\begin{equation}
\forall_\textbf{e}~ \mathcal{T}^0_{[\pi/2]}     (\textbf{e}) = \emptyset,
\label{eq:Rama}
\end{equation}
i.e., every event is X-causally connected to any other.

\subsection{Oblivious embedding of $\mathfrak{S}$ in an SRX}

To  show  that  superluminal  $S_\lambda$ poses  no  problem,  we  now
construct an extension for singlet statistics by the \textit{oblivious
  embedding} of the protocol  $\mathfrak{S}$ in a Newtonian spacetime,
via the procedure given below.  Although the locus of the embedding is
not  relativistic spacetime,  the reproduced  operational correlations
will  be  seen  to  be   non-signaling  and  consistent  with  Lorentz
covariance.

In the  PRF, Alice's and  Bob's measurement events are  denoted $(t_A,
\textbf{x}_A)$ and  $(t_B,\textbf{x}_B)$, respectively.   Without loss
of generality, let $t_B\ge t_A$.
\begin{description}
\item[Pre-sharing  $\chi$  and  $\chi^+$]  The  resources  $\chi$  and
  $\chi^+$ are  pre-shared along the  worldline $W$ used to  share the
  physical, entangled particles.
\item[Free will] Alice  and Bob choose their  inputs freely, according
  to definition (\ref{defn:AS}), with the  scope of the past being the
  past half  in the PRF. Thus  the concept of free  will is manifestly
  Lorentz non-covariant.
\item[Superluminal   transmission  of   $\mathcal{R}$]  The   resource
  $\mathcal{R}$ is  transmitted from  Alice to  Bob at  infinite speed
  ($v_\lambda  = \infty$),  as  seen  in the  PRF.  This ensures  that
  $\mathcal{R}$ reaches Bob in time  for him to output the appropriate
  $y$.
\item[Obliviousness  of   $\chi,  \chi^+,  \mathcal{R}$]   To  enforce
  operational no-signaling, Bob can only  access the final outcome $y$
  directly, but never  $\chi, \chi^+$ and $\mathcal{R}$,  except as he
  may infer by knowing $y$.
\end{description}
Evidently the  above embedding implements  an extension of QM  that is
Lorentz  non-covariant.   Yet  it  is  valid in  the  sense  of  being
predictively equivalent to QM, provided the protocol $\mathfrak{S}$ is
sound. The extension so constructed respects operational no-signaling,
Eq.   (\ref{eq:opnosig}), and  also spontaneity  (\ref{defn:spontan}).
Moreover, Alice and  Bob are also free-willed according  to the stated
(non-covariant) scope  and definition (\ref{defn:AS}).  The  role that
Bob's obliviousness  plays is  crucial, since  otherwise Bob  would be
able to gain some information about  $a$ knowing $y$ and would thereby
in    general    receive    a    superluminal    signal.

In particular,  if Bob could  access $\mathcal{R}$, he would  know $a$
superluminally.  Likewise,  if he  knew $\chi$,  then knowing  $b$ and
obtaining   $y$,  he   would   obtain  some   information  about   $a$
superluminally. For  example, suppose  Alice and Bob  share a  PR box,
which is realized by an equal  mixture of underlying states $\lambda =
\textbf{d}^{0_1}$   and   $\lambda   =  \textbf{d}^{3_1}$   in   Table
\ref{tab:bm1}.  If Bob knows  that at a given time the  box in a given
instance  is  $\textbf{d}^{0_1}$,  and furthermore  setting  $b=0$  he
obtains outcome $y=1$, then he  superluminally knows that $a=1$.  This
obliviousness  of   $\chi$  and   $\mathcal{R}$,  together   with  the
assumption $v_\lambda = \infty$ (see  below), ensures that there is no
superluminal signaling  at the  operational level.   The obliviousness
also means that  $\mathcal{R}, \chi, \chi^+$ take on the  role of HV's
in the derived extension.

Let $v_{\rm  exp} \equiv |\frac{\Delta \textbf{x}}{\Delta  t}|$, where
$\Delta  \textbf{x} \equiv  \textbf{x}_B-\textbf{x}_A$  and $\Delta  t
\equiv t_B-t_A$.  One  can-- in place of a Newtonian  spacetime as the
locus  of the  embedding-- employ  an intermediate  SRX, in  which the
ontological   signals  $S_\lambda$,   transmitted   at  finite   speed
$v_\lambda$  as referred  to the  PRF, propagate  faster than  $v_{\rm
  exp}$ but  not infinitely  so.  It is  straightforward to  adapt the
above embedding procedure for maximal free  will in a Newtonian SRX to
a  procedure  for  embedding  protocol $\mathfrak{S}$  in  a  generic,
possibly non-Newtonian  SRX, with $c<v_\lambda<\infty$.  The  basic idea
is  given   in  Figure  \ref{fig:SRX},  with   details  in  supplement
\ref{xsec:nonNewton}).  Such  an embedding realizes a  valid extension
for QM provided
\begin{equation}
v_\lambda \ge v_{\rm exp}
\label{eq:A1}
\end{equation}
For completeness,  a more general  embedding of a modified  version of
$\mathfrak{S}$, which effects  a reduction in free  will, is presented
in Supplement \ref{xsec:nonNewtonX}).

\begin{figure}
\includegraphics[width=8cm]{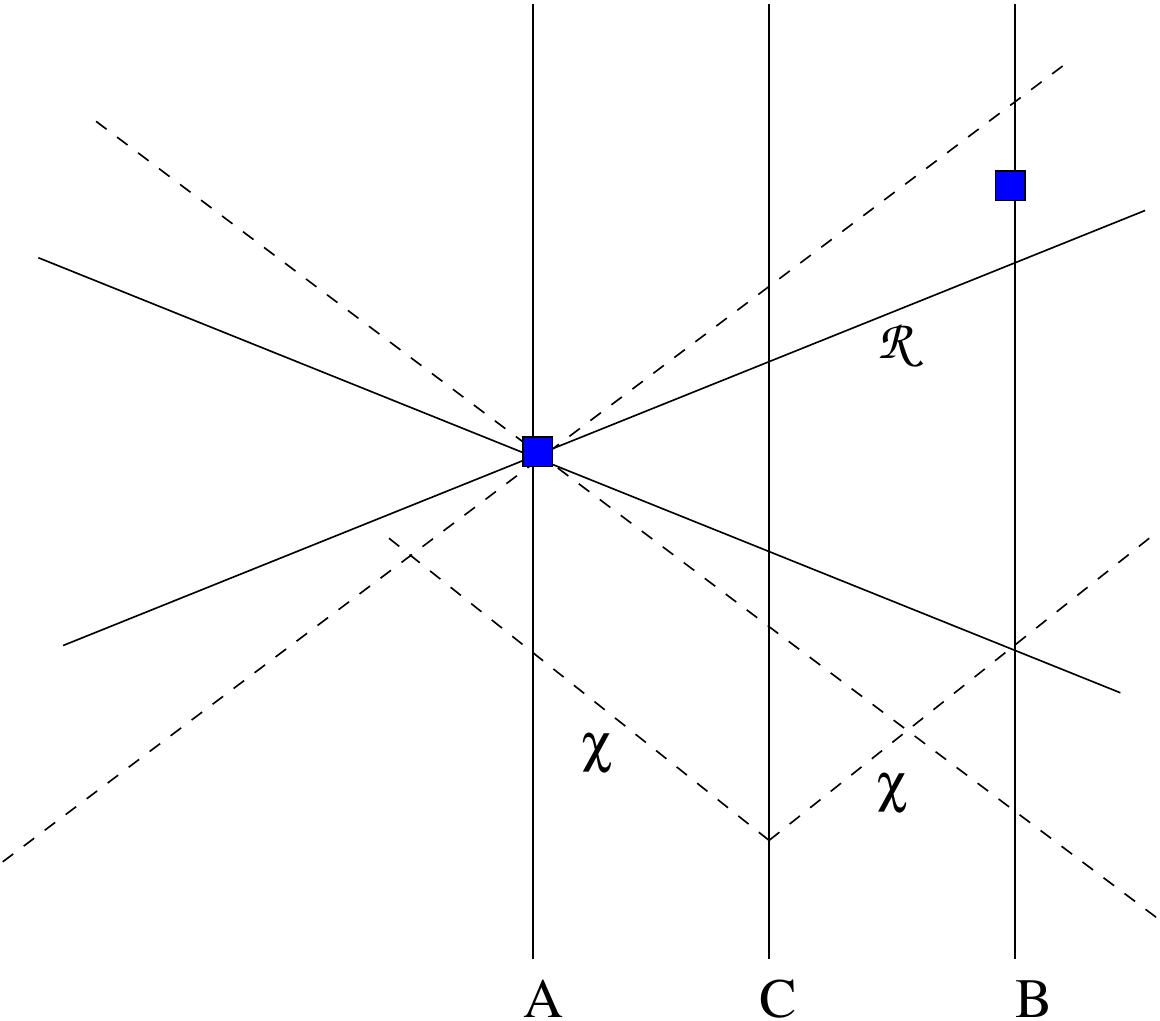}
\caption{\textbf{Oblivious embedding  of protocol $\mathfrak{S}$  in a
    generic  SRX characterized  by $c<v_\lambda\le\infty$}.   Alice's,
  Bob's and  Charlie's worldlines are  the vertical lines  labelled by
  $A, B$ and $C$, respectively.   Alice's and Bob's measurement events
  $\textbf{e}_A$ and $\textbf{e}_B$ are  indicated by the filled (blue
  online) square boxes.  They  are spacelike separated, but X-causally
  connected, with the  X-cones indicated by the  solid slanting lines.
  The lightcone  is indicated by the  dashed lines.  The free  will of
  Alice's and Bob's action  is characterized by (\ref{defn:AS}), where
  ``past'' is the past-half as seen  in the PRF. Randomness $\chi$ and
  $\chi^+$ are pre-shared when the  particles are distributed to Alice
  and Bob, while $\mathcal{R}$ is transmitted from Alice's measurement
  event to Bob's worldline with speed $v_{\lambda}$.}
\label{fig:SRX}
\end{figure}

For  any fixed  SRX  with  $v_\lambda <  \infty$,  one  can choose  an
experimental  set-up such  that Eq.  (\ref{eq:A1}) is  violated, which
would lead  to a  breakdown in quantum  correlations.  Not  only that,
such breakdowns could be used as a basis for superluminal signaling at
the  operational  level  through  the violation  of  hidden  influence
inequalities  \cite{BPA+12,BBL+0}, as  discussed below,  implying that
only  the  Newtonian  SRX, characterized  by  $v_\lambda:=\infty$,  is
unconditionally valid.

\subsection{Hidden influence inequalities and the Newtonian SRX
\label{sec:HIE}}

In our framework, the validity of the extension requires condition the
satisfaction of Ineq. (\ref{eq:A1}).  If it fails (e.g., because Alice
and Bob measured almost simultaneously as seen in the PRF), then there
is a breakdown  in the nonlocal correlations, and the  extension of QM
produced  by  the  embedding  is invalid,  even  though  the  protocol
$\mathfrak{S}$ is sound.

This breakdown  would be  experimentally testable  (cf.  \cite{SBS+14}
and references therein).   But the main difficulty  with the breakdown
is   that   one   can    construct   hidden   influence   inequalities
\cite{BPA+12,BBL+0}  which   can  exploit   this  breakdown   to  send
superluminal signals  \textit{at the operational level},  in violation
of Eq.  (\ref{eq:opnosig}) even when  Alice and Bob  make measurements
freely.

The basic idea here may  explained as follows \cite{Ryf09}.  Alice and
Bob  share  the  state  $\frac{1}{\sqrt{2}}(|0\rangle_A|00\rangle_B  +
|1\rangle_A|11\rangle_B)$, in such a way  that Bob's two particles are
spatially separated from each other by distance $R$, while at the same
time they are equidistant from Alice's particle at distance $L$, where
$L\gg R$.   Alice has  the choice  of measuring  her particle  at time
$t_A$ in  the computational basis  or not  measuring at all.   Bob has
pre-programmed his particles to be measured in the computational basis
simultaneously   at  time   $t_B$.   Let   $\infty  >   v_{\lambda}  >
\frac{L}{t_B-t_A}  >  c$.  Bob's  two  particles  will fail  condition
(\ref{eq:A1}) because  in this case $v_{\rm  exp}=\infty$.  Therefore,
his two particles will produce  uncorrelated outcomes if Alice did not
measure.     But     if    she    does    measure,     then    $v_{\rm
  exp}=\frac{L}{t_B-t_A}$ and  her influence  satisfies (\ref{eq:A1}),
and therefore both  of Bob's particles will be set  to the same value,
producing correlated outcomes.  Thus she can signal him superluminally
if $v_{\lambda}$ is finite.

Therefore, the  only SRX that can  result in a valid  extension is the
Newtonian    SRX,    for    which    $\theta_{\lambda}=\frac{\pi}{2}$,
guaranteeing the general satisfaction of Eq.  (\ref{eq:A1}).  This can
be  shown to  be true,  even going  beyond the  above single-parameter
family of SRXs.   For example, consider the SRX  obtained by replacing
the universal  PRF considered above by  a PRF that is  identified with
the  reference frame  of  Alice's detector  or  Bob's detector.   This
yields a  valid extension  provided Alice  and Bob  are agreed  on the
time-ordering of  their respective detection events  \cite{SS97}.  But
when the  relative motion between the  two detectors is such  that the
two  reference  frames  disagree  on  the  time-ordering  of  the  two
detection events, then a breakdown in the correlations is predicted at
the operational  level, which can be  used as the basis  for violating
operational no-signaling assuming maximal free will \cite{Sua0}.

We conclude  that given  a sound  simulation protocol,  it leads  to a
valid  extension  if and  only  if  the  embedding is  Newtonian.   In
practice,  we can  replace the  infinite value  of $v_{\lambda}$  by a
sufficiently high speed so that  $\mathcal{R}$ traverses the length of
the universe (about $13 \times 10^{10}$ Lyr) in unit time (say, Planck
time,  about $5  \times  10^{-44}$ sec),  which  gives $v_{\lambda}  >
10^{61}c$ \cite{Gar02}.  But  this is a quantum gravity  issue that we
ignore here.  

\section{Implications for ontological extensions
\label{sec:XSIC}}

The embedding of $\mathfrak{S}$ directly demonstrates how to create an
ontological extension for  (the considered fragment of)  QM. Thus, the
simulation resources  can now  be re-interpreted  as variables  in the
ontological extension,  as per Eq.  (\ref{eq:conversion}).  Randomness
$I_R$  and  signaling  $S_R$  can be  replaced  by  the  corresponding
ontological   quantities,   namely   indeterminism   $I_\lambda$   and
ontological signaling $S_\lambda$.

\subsection{Complementarity in the ontological extension}

For simulating a  singlet, we require the resource given  by a $C$-box
with $C=1$.  Substituting this in Eq.  (\ref{eq:main}) yields
\begin{equation}
S_R + 2I_R  \ge 3F-2,
\label{eq:junior}
\end{equation}
where $\frac{2}{3} \le  F \le 1$.  Setting  $F=1$ in (\ref{eq:junior})
gives Alice  and Bob  full free  will in the  sense analogous  to that
considered   in   Refs.    \cite{CK06,*CK09,CR11},    and   gives   us   Eq.
(\ref{eq:senior}) applied to simulation resources: $S_R + 2I_R\ge1$.

Under  the identification  (\ref{eq:conversion}), the  complementarity
(\ref{eq:junior})     becomes     the    corresponding     ontological
complementarity: $S_\lambda + 2I_\lambda  \ge 3F-2$, where the reduced
free  will in  the extension  is implemented  as described  in Section
\ref{xsec:nonNewtonX}.  We  then have  Eq. (\ref{eq:2comp}),  which is
repeated here:
\begin{equation}
S_\lambda + 2I_\lambda \ge 1,
\label{eq:embcomp}
\end{equation}
assuming full free will.  

Result (\ref{eq:embcomp})  can now be  used as  a basis to  derive the
Free  Will   Theorem  \cite{CK06,*CK09}  and  the   Unextendability  Theorem
\cite{CR11}   in   the   context    of   singlet   statistics.    From
Eq. (\ref{eq:embcomp}), it follows that
\begin{equation}
I_\lambda=0 \implies S_\lambda>0,
\label{eq:fwt}
\end{equation}
which is  an operational form  of Bell's theorem.   Eq. (\ref{eq:fwt})
asserts  that  any  deterministic  model of  singlet  statistics  must
necessarily  be  signaling.  In  the  context  of singlet  statistics,
Eq. (\ref{eq:fwt}) is  also the essential mathematical  content of the
Free Will Theorem \cite{CK06,*CK09},  which assumes that $S_\lambda=0$
by  appeal to  SR,  and thereby  concludes  that $I_\lambda>0$,  i.e.,
``particles have free will''.

Further, from (\ref{eq:embcomp}), we have the stronger result:
\begin{equation}
I_\lambda<\frac{1}{2} \implies S_\lambda>0,
\label{eq:ut}
\end{equation}
which  asserts  that  any  predictively  superior  extension  for  the
statistics of singlets  will be signaling.  In the  context of singlet
statistics, (\ref{eq:ut}) is the essential mathematical content of the
unextendability  result  \cite{CR11}.  By  appeal  to  SR and  to  the
definition of  free will (\ref{defn:AS})  with scope of past  given by
$\overline{\mathcal{T}^+}$,  Ref.   \cite{CR11}  also   requires  that
$S_\lambda=0$, and therefore, on basis of Eq.  (\ref{eq:ut}), excludes
predictively superior ($I_\lambda<\frac{1}{2}$) extensions.

Despite this,  our explicit construction of  a non-covariant extension
for QM  showed that  non-vanishing $S_\lambda$ is  ``harmless'', i.e.,
that does not violate operational  no-signaling and does not undermine
a  suitably  defined  free  will.    In  fact,  it  is  necessary  for
constructing predictively  superior extensions.  In this  light, it is
clear  that   the  requirement  that   extensions  of  QM   should  be
non-signaling is unfavorable to extend QM.  The fundamental assumption
of Refs.  \cite{CK06,*CK09,CR11}, that leads them  to this requirement
is that the causal structure of the spacetime of the extension also is
Minkowskian.  The  non-covariance of  an ontological extension  for QM
carries  no  physical  consequence   and  thus  bears  no  operational
significance.  

 We  note that the  complementarity relation we obtained  and the
  above  conclusions   drawn  from   it  would   apply  also   to  any
  non-signaling  operational  theory,  including   one  in  which  the
  CHSH-Bell  inequality  is  violated  up to  the  algebraic  maximum.
  However, the question of why QM  does not allow the violation of the
  CHSH-Bell  inequality up  to its  algebraic maximum  \cite{PR94}, an
  open problem in quantum foundations, is as such not addressed in our
  model.  

\subsection{Bohmian and GRW-type collapse models}

Our  above   analysis  of  the   randomness-signaling  complementarity
transferred to  the obliviously  embedded protocol provides  a general
clarification regarding why there is  no bar against the compatibility
between  SR and  predictively superior  ontological extensions  of QM.
Bohmian mechanics  \cite{Boh52a,*Boh52b} and GRW-like  collapse models
\cite{GRW86,Tum06a,*Tum06b}  provide  specific   instances  where  the
non-covariant ontological  elements are seen to  reproduce a covariant
operational  theory  (which is  exactly  QM  in  the case  of  Bohmian
mechanics).  

 For ontological models derived by embedding simulation protocols
  in  spacetime,   our  approach  shows  that   predictively  superior
  extensions of QM will contain elements in the ontological level that
  are necessarily  non-covariant but ``harmless''.  If  an ontological
  model is not manifestly reducible  to a simulation protocol embedded
  in spacetime in the  above fashion, then the operational/ontological
  level separation  may not  correspond to  the covariant/noncovariant
  division  of elements  in  the  model.  Indeed,  for  the GRW  model
  \cite{Sua10,Tum06a,*Tum06b,BDG+14} and  Bohmian model \cite{DGN+14},
  elements that are recognized as  ontological in the respective model
  are given  a covariant description.   However, it is known  that for
  any  model  of  quantum  nonlocality,  there  would  be  fundamental
  influences  and  fundamental  correlations  that  lack  a  covariant
  description \cite{Har92,Gis11,AA84,Ghi00}, and  this idea receives a
  particularly clear and simple elucidation in our approach.
 
For example,  in the  case of the  Bohmian mechanics,  the information
about  the measurement-induced  deformation of  the quantum  potential
requires instantaneous  signaling in a universal  PRF \cite{BH93}, and
may be identified  with the ontological version  of $\mathcal{R}$ with
$S_\lambda>0$. The  concept of  obliviousness in the  present context,
then, is  analogous to that  of ``absolute uncertainty''  discussed in
Ref. \cite{DGZ92}.

\section{Discussions\label{sec:conclu}}

We  now  briefly  indicate  other   implications  of  our  work.   Our
ontological model of QM derived  from a simulation protocol provides a
``behind the scenes'' mechanism in  the spirit of Bell \cite{DB89} for
explaining quantum  correlations, which (again in  Bell's words) ``cry
out for  explanation'' \cite[Ch.  \ 9]{Bel87}.  Moreover,  without the
ontological  extendability of  SR,  the experimental  fact of  quantum
nonlocality would  compel us to  regard free will and  no-signaling as
logically dependent. We saw that in the Newtonian extension, free will
can  coexist with  superluminal signaling.   The concept  of SRX  thus
frees us from this epistemological obligation.

QM and  relativity theory  form the  corner-stones of  modern physics.
Yet,  ironically, both  have  resisted unification.   It is  generally
acknowledged that the reasons for  this impasse are related to general
relativity,  and that  quantum field  theory evidences  the harmonious
unification  of QM  and SR.   However, studies  in the  foundations of
quantum  nonlocality  suggest  that  there is  a  ``tension''  between
quantum  nonlocality  and  SR  in  the  sense,  as  seen  above,  that
non-trivial extensions of QM will be  signaling.  Yet we saw that such
extensions  need  not  pose  a  threat  either  to  free  will  or  to
operational no-signaling.   On this  strength, we  are led  to believe
that the unification of QM  with general relativity in quantum gravity
may also profit from a similar  exercise, namely to unify the theories
by unifying suitable ontological extensions.

\begin{acknowledgments} 
SA acknowledges  support through the INSPIRE  fellowship [IF120025] by
the Department of Science and  Technology, Govt.  of India and Manipal
university graduate programme. RS acknowledge support from the DST for
projects SR/S2/LOP-02/2012.
\end{acknowledgments}

\bibliography{quantarv}

\appendix

\section*{\large Supplemental notes}

\section{Oblivious embedding of protocol $\mathfrak{S}$
in a generic SRX: Free-willed scenario\label{xsec:nonNewton}} 

We consider the case  where Alice and Bob  have maximal free
will, but the  SRX is not necessarily Newtonian.   Suppose Alice's and
Bob's  spacelike-separated measurement  events are  $\textbf{e}_A$ and
$\textbf{e}_B$, respectively.  As  seen in the PRF,  let the spacetime
coordinates  of   these  two  events  be   $(\textbf{x}_A,  t_A)$  and
$(\textbf{x}_B,t_B)$.   Further, let  $W$ denote  the worldline  along
which their respective  particles were received from  a quantum source
(the dashed lines labelled $\chi$ in Figure \ref{fig:SRX}).  We define
the    \textit{oblivious    embedding}    of    simulation    protocol
$\mathfrak{S}(\chi,\mathcal{R})$  in  an   SRX  $\theta_{\lambda}$  as
follows (see Section \ref{sec:srX} and Figure \ref{fig:SRX}).

\begin{description}
\item[Pre-sharing  $\chi$  and  $\chi^+$]  The  resources  $\chi$  and
  $\chi^+$ are pre-shared along spacetime path $W$.
\item[Free will] Alice  and Bob choose their  inputs freely, according
  to definition (\ref{defn:AS}),  with the scope of the  past being he
  past half in the PRF, and  not the complement of the future
  lightcone.    Thus  the   concept   of  free   will  is   manifestly
  non-covariant.
\item[Superluminal  transmission  of $\mathcal{R}$  at  $v_{\lambda}$]
  Without  loss of  generality, let  $t_A <  t_B$ in  the PRF.   It is
  assumed that the two events are X-causally connected, so that
\begin{equation}
\frac{1}{c}\frac{|\textbf{x}_A-\textbf{x}_B|}{|t_A-t_B|}  \le
\tan\left(\theta_{\rm DI}\right),
\label{eq:newton}
\end{equation}
and information about Alice's input  reaches Bob's station in time for
his  measurement at  $\textbf{e}_B$  as seen  in  the PRF.   Condition
(\ref{eq:newton}) also  is manifestly non-covariant.  (The  case where
Eq.    (\ref{eq:newton})  fails   is  considered   below  in   Section
\ref{sec:HIE}).  Together with $\chi$  and $\chi^+$, this suffices for
his  station  to   compute  his  outcome  $y$,   consistent  with  the
predictions of QM, by assumption of soundness of the protocol.
\item[Obliviousness  of  $\chi,  \chi^+, \mathcal{R}$]  Bob  can  only
  access the final outcome $y$  directly, but never $\chi, \chi^+$ and
  $\mathcal{R}$, except as he may infer by knowing $y$.
\end{description}

Under  the   embedding,  the   simulation  resources   $\chi,  \chi^+,
\mathcal{R}$ take on  an ontological significance in  the extension so
defined. The  extension is Lorentz non-covariant,  since $\mathcal{R}$
and free  will are not  covariant, being  defined with reference  to a
PRF.   One  might  say  that  ``there  is  no  story  in  relativistic
spacetime''   of  nonlocality   (cf.   \cite{Gis11}).    By  contrast,
spontaneity being an operational concept, the scope of the past in its
definition, which  is the complement  of the  causal future in  SR, is
covariant.

The non-covariant definition of free  will protects free will from the
threat potentially posed by the ontological superluminal signaling: at
$\textbf{e}_A$, the conditions  (\ref{eq:ontnosig}) forbid ontological
signaling  into the  past half  as seen  in the  PRF, whereas  Alice's
signal  is directed  into  the future  half  as seen  in  the PRF;  at
$\textbf{e}_B$, Bob transmits  no signal anyway, and  hence his action
does  not contradict  (\ref{eq:ontnosig})  in the  stated scope.   The
crucial difference  between the  present definition  of free  will and
that in Ref.  \cite{CR11} is in the scope of the past.

\section{Oblivious embedding of protocol $\mathfrak{S}$
in a generic SRX: Reduced freewill scenario\label{xsec:nonNewtonX}}

 Thus any given $\Lambda$ above 2 and up
to 4 can be achieved by just reducing the free will (\ref{defn:Hal10})
from the maximal value of $F=1$ to
\begin{eqnarray}
F^\prime   &\equiv&   1   -   \frac{1}{3}\left(\frac{\Lambda}{2}-1\right)   =
1-\frac{C_\Lambda}{3},
\label{ineq:F'}
\end{eqnarray}
where  $C_\Lambda\ge0$. 

Let  $\textbf{P}_\mathcal{L}$ be  the ``0-bit  protocol'' obtained  by
uniformly  mixing  the local-deterministic  boxes  $\textbf{d}^{j_0}$.
Denote by  $\textbf{P}_\mathcal{L}^\ast(\Lambda)$ the new  protocol to
realize  $\Lambda$, obtained  via reduction  of free  will applied  to
$\textbf{P}_\mathcal{L}$.     This    requires    that    we    choose
$F{:=}F^\prime$.        It        can       be        shown       that
$\textbf{P}_\mathcal{L}^\ast(\Lambda)  \in  \mathfrak{F}^P$, as  shown
earlier.

A more  intuitive and  implementationally straightforward  approach to
Theorem   \ref{thm:main},  would   be   as   follows.   The   enhanced
$\mathcal{L}$ mode can be visualized as a probabilistic mixture of the
bound mode $\mathcal{L}$ and free  mode $\mathcal{F}$, with both modes
being  set  at  the  observed level  of  violation  $C_\Lambda$.  Mode
$\mathcal{L}$  (resp.,  $\mathcal{F}$)   is  played  with  probability
$p_\mathcal{L}$                 ($p_\mathcal{F}$),                with
$p_\mathcal{F}+p_\mathcal{L}=1$.     The    ``average''   free    will
(henceforth also denoted $F$) in this ``mixed mode'' will be
\begin{eqnarray}
F &=& p_\mathcal{F}\cdot 1  + p_\mathcal{L}\cdot  F^\prime,\nonumber \\
  &=& 1     -
\frac{C_\Lambda}{3}\left(1-p_\mathcal{F}\right),
\label{eq:pF0}
\end{eqnarray} 
using Eq. (\ref{ineq:F'}).  It follows from Eq. (\ref{eq:pF0}) that if
$C_\Lambda>0$, then
\begin{equation}
p_{\mathcal{F}} = 1 - \frac{3}{C_\Lambda}(1-F).
\label{eq:pF}
\end{equation}
If $C_\Lambda=0$,  then we  set $F=1$ irrespective  of $p_\mathcal{F}$
according to  Eq.  (\ref{eq:pF0}).   Averaging over  the communication
costs in the two scenarios, we find
\begin{eqnarray}
S_R + 2I_R &\ge& p_{\mathcal{F}}C_\Lambda,
\label{eq:3main}
\end{eqnarray} 
and Eq.  (\ref{eq:main}) then follows using Eq.  (\ref{eq:pF}).

Therefore,  an  implementation  to  simulate  the  violation  of  CHSH
inequality at  a given level  $\Lambda$ would be  by probabilistically
mixing   a    protocol   $\textbf{P}^\ast_\mathcal{L}(\Lambda)$   with
$\textbf{P}_\mathcal{F}(\Lambda)$,   which  is   a  $\mathcal{F}$-mode
protocol in $\mathfrak{F}^P$ that  violates (\ref{eq:bi}) to the level
$\Lambda$. The required mixed mode protocol is given by
\begin{equation}
\textbf{P}_\mathcal{LF}^\ast(\Lambda)       \equiv       p_\mathcal{L}
\textbf{P}^\ast_\mathcal{L}(\Lambda)          +          p_\mathcal{F}
\textbf{P}_\mathcal{F}(\Lambda).
\label{eq:mixed}
\end{equation}  
Let $\chi^\ast$ be a bit string that encodes instructions on realizing
$\textbf{P}_\mathcal{LF}^\ast(\Lambda)$.  One  way to  use $\chi^\ast$
would   be   as  follows:   $\chi^\ast$   carries   two  bit   strings
$\chi^\ast_{1}$      and       $\chi^\ast_{2}$:      one      executes
$\textbf{P}^\ast_\mathcal{L}(\Lambda)$   when   the   $j$th   bit   of
$\chi^\ast_{1}$  (denoted  $\chi^\ast_1(j)$)  is  ``0''  and  executes
$\textbf{P}_\mathcal{F}(\Lambda)$  when  $\chi^\ast_j(j)=1$.  The  bit
string  $\chi$ is  used  to realize  the $\textbf{P}_\mathcal{L}$  and
$\textbf{P}_\mathcal{F}$,  while   $\chi^\ast_2$  is  used   to  boost
$\textbf{P}_\mathcal{L}$ to $\textbf{P}_\mathcal{L}^\ast$.

The           $C$-box           defined          by           protocol
$\textbf{P}_\mathcal{LF}^\ast(\Lambda)$ in Eq. (\ref{eq:mixed}) can be
used as a subroutine to simulate a singlet, which would require mixing
$\textbf{P}^\ast_\mathcal{L}(\Lambda=4)$                           and
$\textbf{P}_\mathcal{F}(\Lambda=4)$          protocols.           Then
$\textbf{P}_\mathcal{LF}^\ast(\Lambda)$ will in general be an ``biased
PR box'' \cite{ASQIC}, i.e., one  that attains $\Lambda=+4$ but may be
signaling  (Section \ref{sec:redu}).   .  We  denote this  $\chi^\ast$
enhanced    protocol   for    simulating    singlet   statistics    by
$\mathfrak{S}(\chi,\chi^+,\chi^\ast,\mathcal{R})$. In such a protocol,
although there is  reduction in free will, there will  be no reduction
in spontaneity, which is a reasonable requirement.

Embedding protocol $\mathfrak{S}$ with reduced free will is similar to
the above situation  with maximal free will,  except that additionally
it requires  pre-sharing or  superluminal transmission of  bit strings
$\chi^\ast$  and   (as  explained  below)   $\mathcal{R}^\ast$.  These
additional  resources must  also  be embedded  obliviously.  We  shall
consider two situations.

The       first       one      involves       embedding       protocol
$\mathfrak{S}(\chi,\chi^+,\chi^\ast,\mathcal{R})$      to      realize
$\textbf{P}_\mathcal{LF}^\ast(\Lambda{:=}4)$, i.e., one  in which free
will can  be reduced by  correlation between the underlying  state and
Alice's and Bob's  choices.  In this case, in addition  to bit strings
$\chi$ and $\chi^+$, the string  $\chi^\ast$ is pre-shared in the same
way, along worldline $W$.  In one  role, $\chi^\ast$ is used to choose
between the free-willed  and bound modes in the mixed  mode picture to
realize  $\textbf{P}^\ast_\mathcal{LF}$.   String  $\chi$ is  used  to
realize    $\textbf{P}_\mathcal{L}$    and    $\textbf{P}_\mathcal{F}$
individually, while  $\chi^\ast$ in its  second role is used  to boost
$\textbf{P}_\mathcal{L}$ to  $\textbf{P}_\mathcal{L}^\ast$.  This will
realize  a  $C$-box  with  $C=1$.  Finally  this  maximal  $C$-box  is
consumed,  along with  $\chi^+$, to  realize singlet  statistics.  The
resulting correlations  respect spontaneity  (\ref{defn:spontan}), and
consequently operational no-signaling (\ref{eq:opnosig}).

In the second method, which also implements the scenario where Alice's
and Bob's choices  are spontaneous but lack free will,  we shall allow
Alice's choice  to influence Bob's.  We  introduce $\mathcal{R}^\ast$,
which  is a  secondary  resource superluminally  transmitted at  speed
$v_{\lambda}$ from Alice in  the above embedding procedure.  Among many
ways to  use this, a  convenient one  is to let  $\mathcal{R}^\ast$ to
function just  like $\chi^\ast$.   For example suppose  the underlying
state is  $\lambda = \textbf{d}^{0_0}$.   If Alice selects  $a=0$ then
$\mathcal{R}^\ast$ permits Bob  to choose either input,  but if $a=1$,
then $\mathcal{R}^\ast$  instructs him  to preferably choose  $b=0$ in
order to enhance the operationally observed $\Lambda$.

The comprehensive  simulation protocol, available for  embedding, will
be  denoted  $\mathfrak{S}   (\chi,  \chi^+,  \chi^\ast,  \mathcal{R},
\mathcal{R}^\ast)$, which may generally involve using both $\chi^\ast$
and $\mathcal{R}^\ast$.

\end{document}